# Lateral shearing optical diffraction tomography of brain organoid with reduced spatial coherence

Paweł Gocłowski[1], Julianna Winnik[2], Vishesh Dubey[1], Piotr Zdańkowski[2], Maciej Trusiak[2], Ujjwal Neogi[3], Mukesh Varshney[3], Balpreet S. Ahluwalia[1,4 #], Azeem Ahmad[1,*, #]

[1]*Department of Physics and Technology, UiT The Arctic University of Norway, 9037 Tromsø, Norway*
[2]*Warsaw University of Technology, Institute of Micromechanics and Photonics, 8 Sw. A. Boboli St., 02-525 Warsaw, Poland*
[3]*The Systems Virology Lab, Division of Clinical Microbiology, Department of Laboratory Medicine, Karolinska Institutet, Stockholm, Sweden.*
[4]*Department of Clinical Science, Intervention and Technology, Karolinska Institute, Stockholm, Sweden*

**Corresponding author:* ahmadazeem870@gmail.com, balpreet.singh.ahluwalia@uit.no
*Email:* pawel.goclowski@gmail.com

[#]*Shared authors*

**Abstract:** Optical diffraction tomography (ODT) is a powerful technique for quantitative, label-free reconstruction of the three-dimensional refractive index (RI) distribution of biological samples. While ODT is well established for imaging thin, weakly scattering samples, it encounters significant challenges when applied to heterogeneous, strongly scattering thick samples such as tissues and organoids. In this work, a novel common-path interferometric approach to ODT is presented, specifically designed for the RI reconstruction of heterogeneous and highly scattering samples at high temporal stability. The proposed technique, termed lateral shearing (LS)-ODT, incorporates partial lateral shearing off-axis interferometry to suppress the effects of multiple scattering, similar to the mechanism in differential interference contrast (DIC) microscopy, which is widely used for imaging thick specimens. Additionally, the LS-ODT system uses dynamic speckle illumination to enhance both spatial phase and RI sensitivity compared to laser-based ODT systems. The effectiveness of this method is demonstrated through experiments on a cell phantom. Its robustness and accuracy are further validated across a wide range of samples, including mouse kidney tissue sections and brain organoids derived from human induced pluripotent stem cells (iPSCs), in both thin and thick sections. Furthermore, correlative fluorescence and RI tomography of the organoids highlights the potential of LS-ODT to enhance and support a broad spectrum of biomedical studies, particularly in the field of histology.

## 1. Introduction:

Optical diffraction tomography is a powerful imaging technique used to reconstruct the three-dimensional (3D) refractive index (RI) distribution of transparent and semi-transparent samples [1, 2]. ODT combines ideas from optical holography, computational imaging, and inverse scattering theory to deliver label-free, high-resolution imaging. Its previous applications include biomedical imaging of cells and tissues, as well as material science [3, 4]. ODT uses various methods for the reconstruction of 3D-RI distribution of a sample. These methods are broadly characterized according to their illumination methods, phase recovery techniques, and 3D reconstruction procedures [5]. ODT systems may generally be categorized into two main types: interferometric-based ODT [6] and non-interferometric ODT [7], or else referred to as intensity-based ODT methods.

Interferometric-based ODT relies on the utilization of coherent light sources and interferometry for the recording of the scattered field's amplitude and phase. The most common and well-established technique inside this category is digital holographic microscopy (DHM), which can be implemented in both common-path and non-common-path interferometry modes. These methods utilize reference beams and interference patterns to reconstruct optical phase information. For non-common-path interferometry, the object and reference beams travel along different paths through the optical system, which compromises the temporal phase stability due to independent fluctuation in the path lengths of the interferometer. In contrast, common-path interferometry improves phase stability by making the sample and reference beams share a common optical path, minimizing the system's sensitivity to the environmental and mechanical noises. The phase retrieval algorithm of DHM depends on the type of recorded interferogram, which is determined by the optical setup - either off-axis or on-axis.

Off-axis holography records interference fringes with a tilt in the reference beam, which allows for the separation of Fourier peaks in the Fourier domain and single-shot phase recovery of the specimen. On-axis holography records interference fringes with either a collinear reference beam or a slightly tilted reference beam with respect to the object beam. Multi-frame phase retrieval procedures are required in this configuration to extract phase information precisely. So far, most of the ODT systems are based on non-common-path off-axis holography optical configurations, such as the Mach-Zehnder interferometer, and typically use high-coherent laser sources [6, 8, 9]. These systems are inherently prone to temporal instability because of their non-common-path nature. Thus, they are susceptible to phase drifts caused by environmental disturbances, such as vibrations or air turbulence [10]. Note that temporal instability has an even greater impact in ODT than in 2D phase microscopy, as the acquisition of multiple images from a large set of illumination angles requires more time, resulting in reduced accuracy.

In addition, most of the non-common-path off-axis systems require a highly coherent illumination for interferogram formation. Such coherent illumination reduces spatial phase sensitivity, due to speckle noise and coherent artifacts that further deteriorates the image quality. To improve both the temporal stability and the phase sensitivity of ODT systems, various alternative optical configurations have been proposed. These include the Wollaston prism-based arrangement, diffraction phase microscopy (DPM), polarization diffraction gratings, and quadriwave lateral shearing interferometry (QWLSI), all utilizing common-path interferometry [11-14]. ODT systems with Wollaston prisms and polarization diffraction gratings operate in total shear mode, i.e., they require sparse samples to provide a sample-free reference beam to be used for interferogram formation. Such configurations are therefore challenging to use for imaging biological samples with high confluency, e.g., organoid and tissues, where it is difficult to find a sample-free region. On the other hand, DPM-based ODT systems generate a reference beam through low-pass filtering the sample information through a pinhole. In such grating-based ODT systems, it is necessary that a sufficiently high grating pitch is employed to diffract the spatial information into multiple diffraction orders without any loss of the spatial resolution. As a result, an additional magnification of 2.5× in the 4f geometry is typically required to satisfy the Nyquist criterion at the camera for lossless recording of the spatial information [12]. QWLSI-based ODT systems can work with low-coherent light sources, resulting in high RI and spatial phase sensitivity. However, one of the main limitations of QWLSI, similar to DPM-based ODT, is its relatively small field of view (FOV). In QWLSI, a high magnification (typically 100×/1.3 NA) is used before the grating to ensure proper sampling of the object information. This allows the diffraction grating to split the spatial details into multiple diffraction orders without any loss of spatial resolution. This increased magnification results in a reduced FOV, thereby limiting the region that can be imaged.

The non-interferometric or intensity-based ODT methods reconstruct the RI distribution based on intensity-only measurements (images) and do not require a reference beam [15, 16]. These methods generally rely on numerical phase retrieval algorithms, such as the transport of intensity equation (TIE) [16] and ptychographic phase retrieval [17]. While these setups simplify experiments and offer improved immunity to the environmental noise, they rely heavily on iterative algorithms, which are computationally expensive and require a large number of intensity images for accurate reconstruction. Additionally, since the phase information in intensity-based ODT is not directly measured but retrieved from the intensity images, its accuracy and sensitivity are usually lower than in the interferometric-based ODT.

Despite the limitations of conventional ODT systems, it has proven to be a valuable tool in the life sciences. However, their application has been largely limited to imaging thin biological specimens, typically less than 10 – 20 µm in thickness. These systems face significant challenges when applied to thick, strongly scattering biological samples. This is due to the multiple scattering effects in thick specimens that degrade the reconstruction quality as most of the reconstruction algorithms assume a weekly scattering specimen. Early implementations of ODT utilized weak scattering approximations such as the Born and Rytov approximations and were primarily utilized in low RI contrast specimens. In Born approximation-based ODT, the scattered field is modeled as a linear function of the scattering contrast (e.g. refractive index). The process is computationally simple and resource efficient, but the technique cannot make an accurate reconstruction of highly scattering samples. In the first order Rytov approximation, the scattered field is modeled as an exponentially dependent function of the scattering contrast and is therefore more suitable for biological samples. However, it still does not work well for highly scattering specimens such as thick organoids or tissue sections, when the strong multiple-scattering effects contaminate the measured phase. Another challenge of a thick specimen is that it decorrelates the sample and the reference fields, which prevents the formation of interference fringes. This decorrelation arises from multiple scattering events within the sample, which distort the phase and amplitude of the transmitted wavefront, making accurate reconstruction of the RI distribution difficult.

To address these limitations, recent developments have explored low-coherence, intensity-based holotomography systems, which are capable of imaging relatively thick specimens (~100 µm) [18]. In parallel, new multi-scattering ODT algorithms have emerged to more accurately model complex light–matter interactions in thick media. The most advanced algorithms in this category are full-wave inversion [19] and modified Born-series ODT [20], which employ rigorous models of inverse scattering, e.g., iterative optimization algorithms, to recover from the multiple scattering effects. While these algorithms provide the most accurate result for highly scattering media, they are computationally expensive and require significant memory resources. The other recent development in ODT is the use of deep learning-based RI reconstruction. Deep learning techniques have been applied to accelerate phase retrieval and RI reconstruction, where neural networks are trained on simulated or experimental data to enhance precision and reduce noise [21-24]. These methods result in faster reconstruction with improved robustness at the cost of needing large datasets and intensive computation.

The growing interest in developing ODT solutions that address the challenge of imaging highly scattering and heterogeneous samples is largely driven by rapid advancements in biomedical research, particularly in the area of organoids. In the following section, the importance of organoids is briefly discussed, with a special focus on brain organoids and their unique imaging requirements.

**Organoids and their imaging needs**

Organoids represent a significant advancement in biomedical research, acting as miniature, simplified versions of human organs that capture the complex architecture, cellular composition, and dynamic developmental processes of native tissues. Derived from pluripotent stem cells, these three-dimensional (3D) cultures have become invaluable for in vitro studies of development, disease modeling, and drug discovery by offering human-relevant systems that bridge the gap between traditional 2D cultures, animal studies and human biology [25-30]. Among these, brain organoids, generated from human induced pluripotent stem cells (iPSCs), have gained prominence for their ability to mimic early human neurodevelopment, including progenitor expansion, cortical layer formation, and neuronal differentiation. They contain a diversity of neural cell types, from radial glial stem cells to mature neurons, and exhibit spatially stratified compartments that closely resemble fetal brain organization [31, 32]. As such, brain organoids are widely used to study disorders such as epilepsy, autism, and Alzheimer's disease [33, 34]. However, as organoids grow in size and complexity, they become increasingly challenging to image using conventional techniques. Confocal, two-photon, and light-sheet microscopy provide molecular specificity but are limited by photobleaching, optical penetration depth, and the need for exogenous labeling [35, 36] cytoarchitecture, especially in mature organoids that may exceed 100 µm in thickness. These limitations create an urgent need for non-destructive, label-free volumetric imaging methods that preserve structural fidelity while offering cellular-level insight. These limitations become especially apparent when attempting to correlate spatial structure with cell-type distribution across developmental stages.

To overcome these barriers, LS-ODT imaging modalities provide a compelling alternative. By reconstructing the three-dimensional RI distribution of intact organoids, ODT enables the visualization of intrinsic structural contrast, capturing features such as nuclear density, neuropil architecture, and boundary zones between progenitor-rich and neuron-dense regions. Importantly, correlative imaging combining RI tomography with fluorescence microscopy enables comprehensive mapping of both molecular identity and biophysical morphology without requiring physical sectioning, laying the foundation of label-free organoid visualization and analysis.

**Concept study - comparison of three interferometric arrangements of ODT**

Figure 1 (A–C) illustrates three interferometric arrangements: non-common-path mode, total shear common-path mode, and partial shear common-path mode (the configuration used in our proposed method). The variation in the fringe contrast with sample thickness is conceptually illustrated for all three optical configurations. In the non-common-path mode and total shear common-path mode (Fig. 1D), fringe contrast decreases as sample thickness increases. In contrast, Figures 1E and 1F highlight the advantages of the partial shear common-path mode, which maintains nearly constant fringe contrast regardless of the sample thickness. In non-common-path mode, the sample and the reference fields become uncorrelated when imaging a thick sample due to alterations in the sample field as it propagates through the scattering medium, leading to a reduction in fringe contrast. Similarly, in total shear common-path mode, although both superimposing fields originate from the same sample beam, their correlation decreases when the shear amount between the fields exceeds the correlation length [37]. However, in partial shear mode, the correlation between the superimposing fields is maintained, ensuring high fringe contrast regardless of sample thickness. Additionally, Fig. 1 schematically demonstrates that coherence reduction can effectively suppress noise and enhance the quality of interference fringes, compare the noise levels in Fig. 1E and 1F. Figure 1 is intended for conceptual purposes only and aims to illustrate the advantages of partial shear over total shear, as well as the benefits of reduced coherence compared to coherent illumination. Both of these approaches are implemented in the proposed LS-ODT system, making it well-suited for imaging thick and strongly scattering samples. The interferograms shown in Fig. 1D and 1F are generated through MATLAB simulations, whereas Fig. 1E presents an actual experimental interferogram acquired using the LS-ODT system under fully coherent illumination and without a sample.

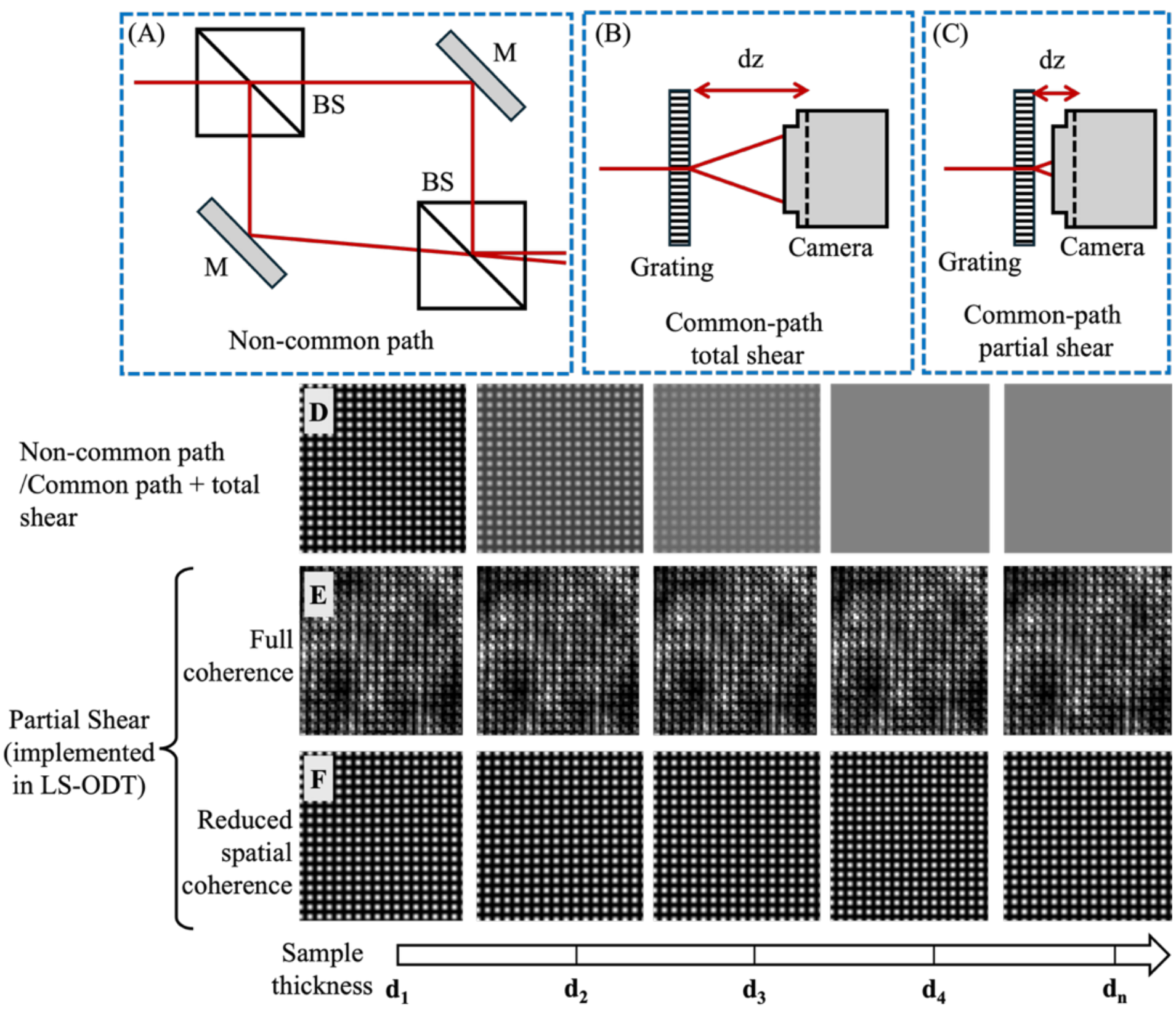

***Figure 1 – Comparison of three interferometric optical configuration. (A–C)** Illustrates three interferometric arrangements: non-common-path mode, total shear common-path mode, and partial shear common-path mode. Demonstration of fringe visibility as a function of penetration depth (d). Synthetic interferograms represent behavior of hypothetical optical system with non-common path geometry or common path geometry with total shear mode **(D)**, partial shear common-path geometry with fully coherent illumination **(E)**, and with reduced spatial coherence **(F)**.*

**Proposed Route: partial coherence common-path lateral shearing (LS)-ODT**

Conventional ODT systems face significant limitations when imaging thick, heterogeneous, and highly scattering biological samples such as organoids, primarily due to multiple scattering and reduced fringe contrast caused by decorrelation between the sample and reference beams. To address these challenges, this work proposes a novel ODT system, partial coherence common-path lateral shearing ODT (LS-ODT), designed to suppress multiple scattering, maintain high temporal stability, and preserve spatial sensitivity. Inspired by the principle of differential interference contrast (DIC) microscopy [38], LS-ODT employs a differential imaging approach that may enhance optical sectioning, making it particularly effective for high-resolution RI tomography of thick specimens where conventional ODT methods fall short.

Figure 2 illustrates the schematic diagram of the LS-ODT system, which incorporates two orthogonally oriented Sagnac interferometers (SI). Each SI consists of a polarizing beam splitter (PBS) and a set of two mirrors. The laser beam is directed toward a rotating diffuser (RD), which reduces spatial coherence while preserving the high temporal coherence of the laser source, thereby generating a dynamic speckle illumination (DSI). With DSI, studies have demonstrated that, despite reduced spatial coherence, high-quality interference fringes can only be observed across the entire field of view (FOV) if the speckle fields in the object and reference arms of the interferometer are mutually correlated [37]. Further details regarding the role of coherence are provided in the Materials and Methods section.

The output of the RD is coupled into a multimode fiber to deliver partially spatially coherent illumination to the illumination module of the tomographic system, which operates with a fixed sample and rotating illumination. Alternatively, the laser beam can be directed toward a single-mode fiber using flip mirror FM to provide coherent illumination. This configuration enables switching between coherent and partially spatially coherent modes for refractive index (RI) tomography and comparative analysis of the role of coherence in the LS-ODT system. The LS-ODT system operates in an inverted transmission mode and is compatible with standard biological imaging platforms, including glass slides and Petri dishes.

The light is then polarized by a linear polarizer ($P_1$) oriented at a 45-degree angle and obliquely directed toward the sample stage using mirrors $M_1$ and $M_2$. Both mirrors are mounted on a single platform attached to a rotational stage, enabling full 360-degree rotation for multi-directional illumination of the specimen. The illumination angle is equal to 45 degrees in air. The sample information is then collected using a 60×/1.2 NA water-immersion objective lens and projected onto the image plane using lens $L_2$. It is then relayed to the camera plane through lenses $L_3$ and $L_4$. Along this relay path, an optical unit consisting of two orthogonally oriented SIs is inserted to generate four identical replicas of the sample. These units function as common-path interferometers, providing high temporal stability to the LS-ODT system.

The sample information is first passed through SI-1 which consists of $PBS_1$, $M_4$, and $M_5$. The $PBS_1$ splits the input beam into two and directs them toward mirrors $M_4$ and $M_5$, which then recombine the beams at the same $PBS_1$, forming a cyclic interferometer. Thus, SI-1 splits the sample information into two orthogonally polarized beams, A and B. The angles of mirrors $M_4$ and $M_5$ control the separation between the beams at the exit of SI-1, thereby adjusting the fringe density and orientation along the horizontal direction.

The output of the SI-1 is passed through the polarizer ($P_2$) oriented at 45° to match the polarization states of both beams, A and B, which is necessary for forming interference fringes. Both the beams are then sent towards the second orthogonally oriented SI-2, which consists of $PBS_2$, $M_6$, and $M_7$. The SI-2 splits the two beams into four, as shown in the inset of Fig. 2. The angles of mirrors $M_6$ and $M_7$ control the separation between the beams at the exit of the SI-2, thereby adjusting the fringe density and orientation along the vertical direction. At the polarizing beam splitter $PBS_2$, beam A is split into beams 1 and 2, while beam B is split into beams 3 and 4. All four beams pass through a polarizer ($P_3$) oriented at 45°, are superimposed at the camera plane using lens $L_4$, and form a mesh grid interferogram.

Lenses $L_3$ and $L_4$ are arranged in a 4f configuration to relay the microscope's image plane to the camera plane. Lens $L_4$ combines all four beams at the image plane. These replicas overlap at the camera plane (positioned slightly offset by a distance dz) with a slight shear (close the diffraction limit of the system) between the wavefronts along both the x- and y-directions. This configuration produces a differential mesh-type interferogram, which, upon post-processing, provides both the amplitude and phase information of the specimen, which are essential for reconstructing its 3D refractive index (RI) map. Further details of the process are provided in the Materials and Methods section. Supplementary Fig. S1 illustrates the working principle of the Sagnac interferometer and explains how fringe density can be adjusted. Supplementary Fig. S2 presents a 2D ray diagram of both Sagnac units along with polarization states. Supplementary Fig. S3 explains calculation of the shear between the wavefronts.

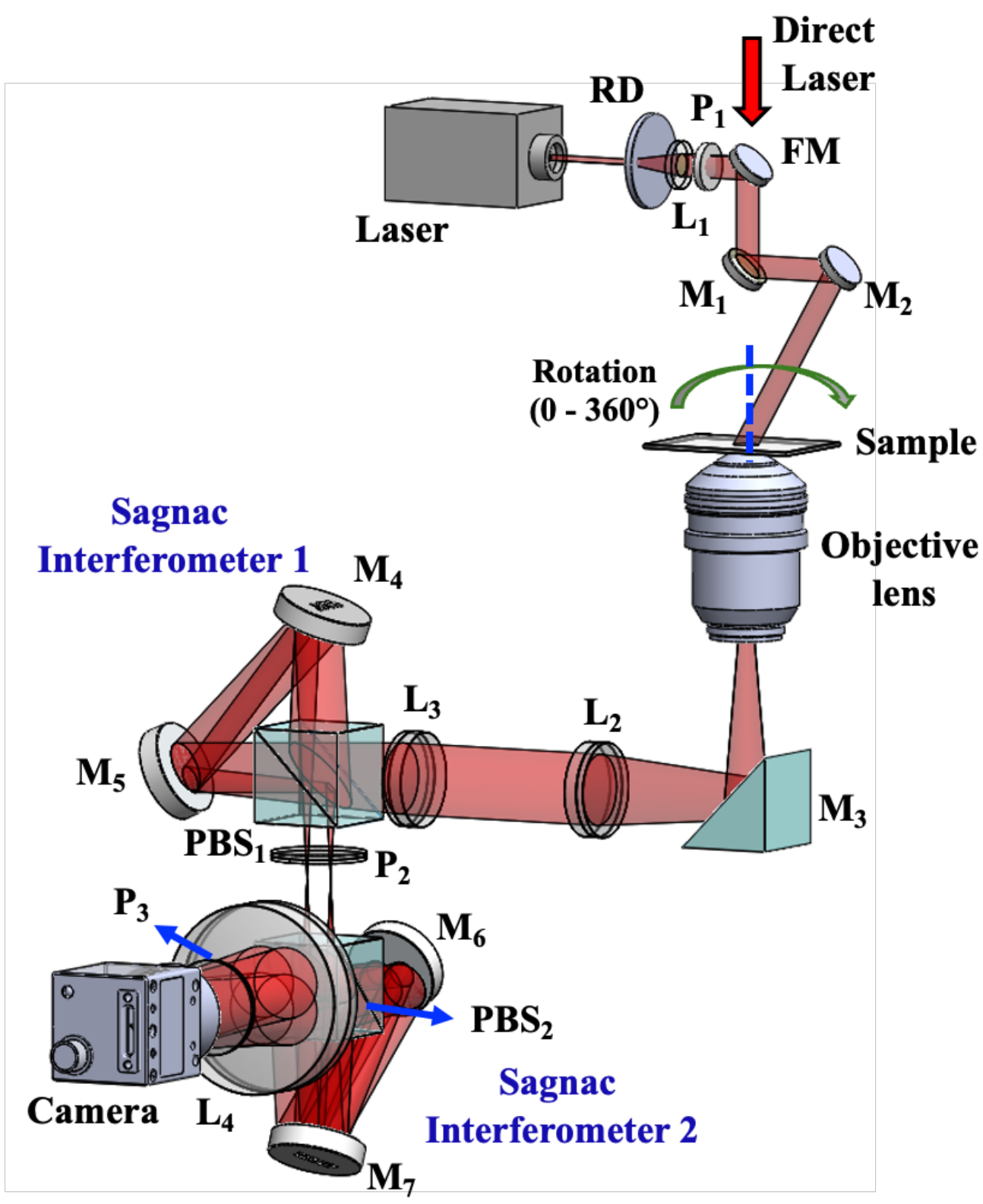

***Figure 2 – Schematic diagram of experimental lateral shearing ODT setup.*** *LS-ODT consists of two orthogonal consecutive Sagnac interferometers. The system uses polarizers ($P_1$ – $P_3$), polarizing beam splitters ($PBS_1$, $PBS_2$), and mirrors ($M_4$-$M_7$), as well as lenses ($L_3$, $L_4$), which generate and recombine four laterally sheared wavefronts, thus producing a mesh-like interferogram that encodes phase gradients along both x and y directions. Scanning illumination is provided by two mirrors ($M_1$, $M_2$) mounted together on a rotational arm. With flip mirror FM it is possible to switch between direct laser illumination and DSI.*

Thus, LS-ODT setup employs a common-path, self-referencing optical configuration that eliminates the need for a separate reference beam to generate interference fringes, unlike conventional ODT systems. This design significantly enhances temporal stability and reduces susceptibility to environmental disturbances such as vibrations or air currents. Additionally, the use of a DSI in LS-ODT provides high phase sensitivity, facilitating accurate RI tomography of the specimens.

The LS-ODT system operates in partial shear mode, which is essential for suppressing multiple-scattering effects. These effects introduce an incoherent background that degrades the accuracy of phase retrieval. A key advantage of the partial shear common-path approach is that all three beams experience the same multiple-scattering effects, allowing these distortions to effectively cancel out during the recording of the mesh grid interferogram. This cancellation significantly enhances the signal-to-noise ratio (SNR) and improves the precision of phase reconstruction. As a result, the LS-ODT system is particularly effective for imaging thick, strongly scattering samples - scenarios in which conventional ODT techniques often face limitations. This makes LS-ODT well-suited for high-sensitivity imaging across a wide range of biological specimens, from optically thin cells to dense tissue samples, combining robustness with high RI sensitivity.

## 2. Results:

### 2.1. LS-ODT based ODT system benchmarking – cell phantom 3D imaging

To benchmark our proposed LS-ODT system, first, we conducted experiments on a 3D-printed cell phantom. The details of the cell phantom can be found in Ref. [39]. The phantom is designed to mimic a real biological cell and contains several internal structures with precisely defined dimensions and refractive indices, including a cell nucleus, spherical nucleoli, and resolution test patterns. The cell phantom is immersed in the immersion oil of RI of 1.46 for imaging. Figure 3(A-F) presents the procedure for field reconstruction of the cell phantom, which is further utilized for the corresponding RI map reconstruction. Here, a DSI light source is used to recover the 3D RI map with high RI sensitivity of the cell phantom, which is then compared with the RI maps obtained under coherent illumination in LS-ODT.

The mesh-type interferogram of the cell phantom and its Fourier spectrum are illustrated in Figs. 3A and 3B. The amplitude image of the cell phantom is obtained by performing zero order filtering of the Fourier spectrum, as shown in Fig. 3C. The gradient phase maps of the cell phantom along the X and Y directions are reconstructed by filtering the +1 order Fourier peaks along the X and Y axes, followed by an inverse Fourier transform, as demonstrated in Figs. 3D and 3E, respectively. The gradient phase maps are further numerically integrated to recover the actual phase map of the phantom, as depicted in Fig. 3F. These integrated phase maps and amplitude images are obtained for different illumination angles ranging from 0° to 360° in steps of 2°, enabling the 3D RI map reconstruction of the phantom (Fig. 3G). The details of the RI map reconstruction are provided in the Materials and Methods section.

The X–Y cross-sections of the RI map at two different z-planes, containing horizontal and vertical resolution bars, are shown in Figs. 3H and 3I, respectively. The corresponding line profiles of these bars are depicted in Figs. 3J and 3K. Similarly, X–Y cross-sections of the RI map at two different z-planes obtained under coherent illumination using LS-ODT are illustrated in Figs. 3L and 3M. The corresponding line profiles of the resolution bars along the horizontal and the vertical directions are shown in Figs. 3N and 3O. These results confirm that partial spatial coherence in LS-ODT significantly enhances imaging quality and improves RI sensitivity. It can be clearly seen in Figs. 3H and 3I that three horizontal and three vertical lines, respectively, are distinctly visible in the case of partial coherence LS-ODT. In contrast, coherent noise in Figs. 3L and 3M obscures the resolution bars in the case of coherent illumination LS-ODT. The line profile of the partial coherence LS-ODT dataset reveals three distinct valleys, corresponding to the bars in the resolution test. These valleys are slightly more pronounced in the vertical bars (Fig. 3K) than in the horizontal bars (Fig. 3J). According to the written structures within the cell phantom, the resolution test includes four bars with widths of 300 nm, 300 nm, 500 nm, and 700 nm [39]. Since three of these bars are distinctly resolved, the resolution of the partial coherence LS-ODT system is estimated to be at least 500 nm, but worse than 300 nm.

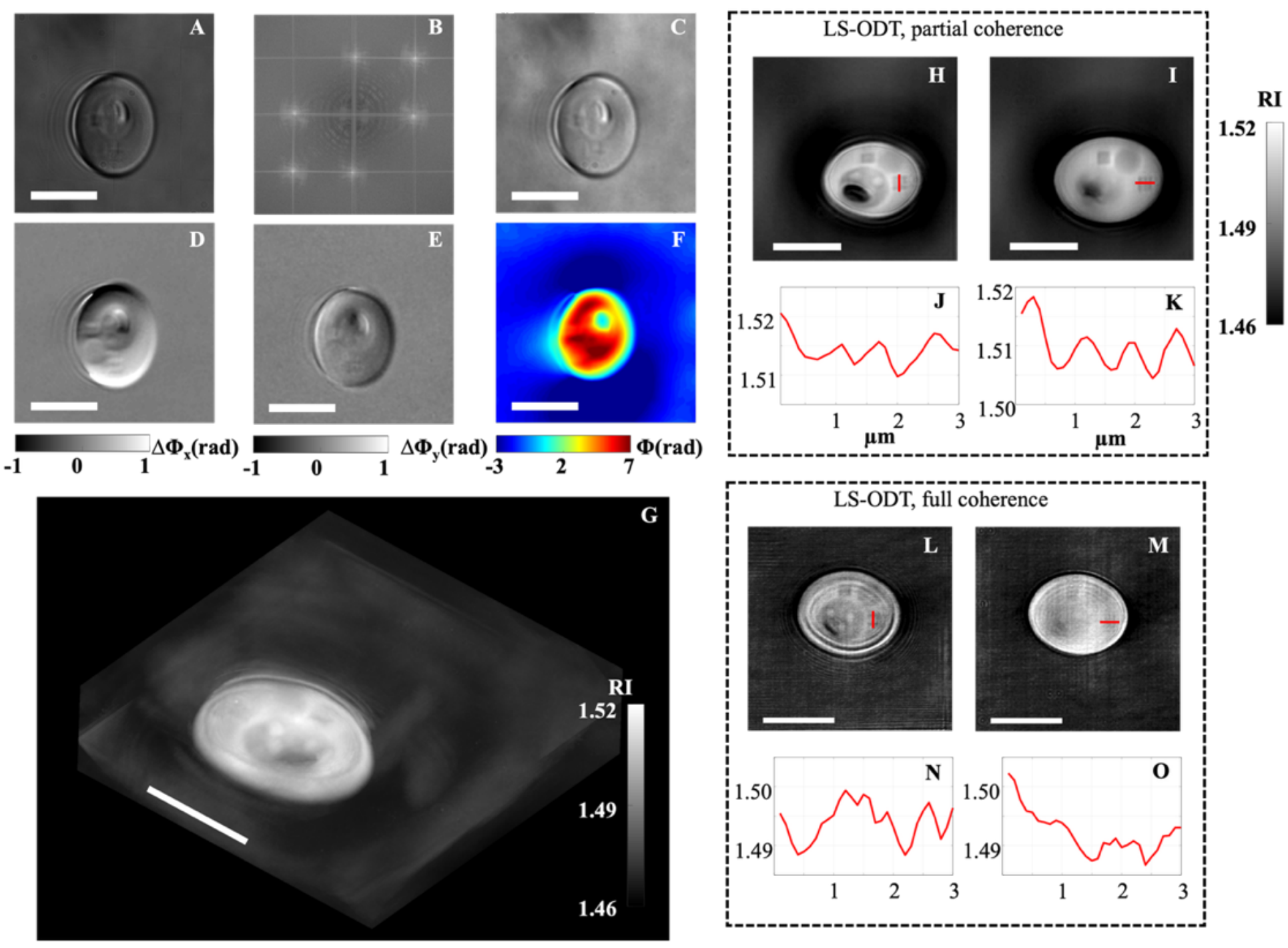

***Figure 3 – Cell phantom imaging.*** *Raw interferogram* ***(A)****, Fourier spectrum* ***(B)****, amplitude* ***(C)****, gradient phase in X-axis* ***(D)****, gradient phase in Y-axis* ***(E)****, integrated phase* ***(F)****, isometric 3D view of RI map* ***(G)****. Two X-Y cross-sections of RI map* ***(H, I)****, with corresponding line profiles, for: LS-ODT partially coherent imaging* ***(J, K)****. Two X-Y cross-sections of RI map* ***(L, M)****, with corresponding line profiles, for: LS-ODT fully coherent imaging* ***(N, O)****. All scale bars correspond to 20 µm.*

In the dataset acquired under coherent illumination LS-ODT, the line profiles exhibit only two dips in both the vertical (Fig. 3O) and horizontal (Fig. 3N) resolution bars. This suggests that the spatial resolution is inferior compared to that of partial coherence LS-ODT, where coherent noise significantly degrades the image quality, leading to the loss of finer structural details. Furthermore, a comparison of the absolute RI values supports the conclusion that the reduced spatial coherence approach enables more accurate RI reconstruction. According to the written structures, the RI of the cell phantom interior is expected to be 1.52, while the RI of the resolution bars should be 1.49. In the case of partial coherence LS-ODT, the RI values lie between 1.51 and 1.52, closely matching the expected values. In contrast, full coherent LS-ODT yields RI values ranging from 1.49 to 1.51, with the resolution bars significantly degraded due to coherent noise.

In case of full coherent LS-ODT, challenge arises at the edges of the phantom, where abrupt RI transitions lead to significant ringing artifacts. However, the partial coherence LS-ODT approach produces reconstructions with fewer artifacts compared to the other methods, owing to the low spatial coherence properties of the DSI light source.

### 2.2. 3D RI imaging of thin bio-specimens

Next, we utilized our method on different biological samples, such as U2OS cell line, thin mouse kidney tissue section (4 µm) and brain organoid section (5 µm) and finally demonstrated the utility for imaging thick brain organoid (120 µm). In addition, we compared the performance of partial coherence LS-ODT with commercial ODT Nanolive system (Nanolive Cell Explorer-Fluo).

The reconstructed 2D and 3D RI map of the U2OS cells is presented in Fig. 4A– 4D. A comprehensive set of intermediate results including the interferogram, Fourier spectrum, amplitude image, gradient phase maps, and integrated phase map is provided in the Supplementary Fig. S4. The 3D RI reconstruction enables visualization of the cells from orthogonal perspectives (Figs. 4B - 4D), although, as expected, the resolution along the axial direction is lower than in the lateral direction. The reconstructed RI values suggest a variation in the range of 1.33 to 1.35, with less abrupt refractive index transitions at the cell boundaries compared to those observed in the cell phantom (see Fig. 3).

Next, the experiment is conducted on a 4 μm thick tissue section of mouse kidney. The RI reconstruction of the tissue section is presented in Figure 4(E-H). The full set of results, including interferogram, Fourier spectrum, amplitude, gradient and integrated phase maps are presented in Supplementary Figure S5. Contrary to the cell line, tissue represents a heterogenous environment of densely packed cells and extra-cellular matrix. Such a complex tissue section from the kidney has been chosen to test robustness of the system in relation to variation in shapes, sizes and RI gradients of the structures. It is observed that partial coherence LS-ODT performed well despite a relatively larger RI variation of 0.07 - 0.08. The refractive index range of the mouse kidney tissue was found to lie between 1.33 and 1.41. It is important to highlight that the sample was hematoxylin-stained, a condition often associated with strong optical absorption. Remarkably, this high absorption did not adversely affect the quality of the reconstructed results, demonstrating the robustness of our approach. The compatibility with hematoxylin-stained samples significantly enhances the practical applicability and versatility of the proposed method, especially considering that hematoxylin staining remains a standard and widely adopted technique in biomedical and histopathological research.

Next, partial coherence LS-ODT was employed to image a 5 μm-thick section of human brain organoids. The resulting 2D and 3D refractive index (RI) maps, including orthogonal views of the reconstructed volume, are shown in Figs. 4(I – L). Intermediate processing steps, such as the raw interferogram, Fourier spectrum, amplitude image, gradient phase maps and integrated phase map are provided in Supplementary Fig. S6. The measured RI values ranged from 1.33 to 1.36, with a total variation of approximately 0.03. Compared to other biological tissues such as mouse kidney (RI range: 1.33–1.41), this narrower RI distribution suggests that brain organoids possess a relatively homogeneous internal structure, consistent with their immature or developing state. This narrow RI contrast likely reflects the dominance of densely packed progenitor cells and early neuronal populations, which have yet to undergo full lamination or myelination [40]. In contrast to the mature tissue, where glial-rich regions and extracellular matrix contribute to higher RI variability, early-stage organoids are typically characterized by compact, radially organized neuroepithelial structures [41, 42].

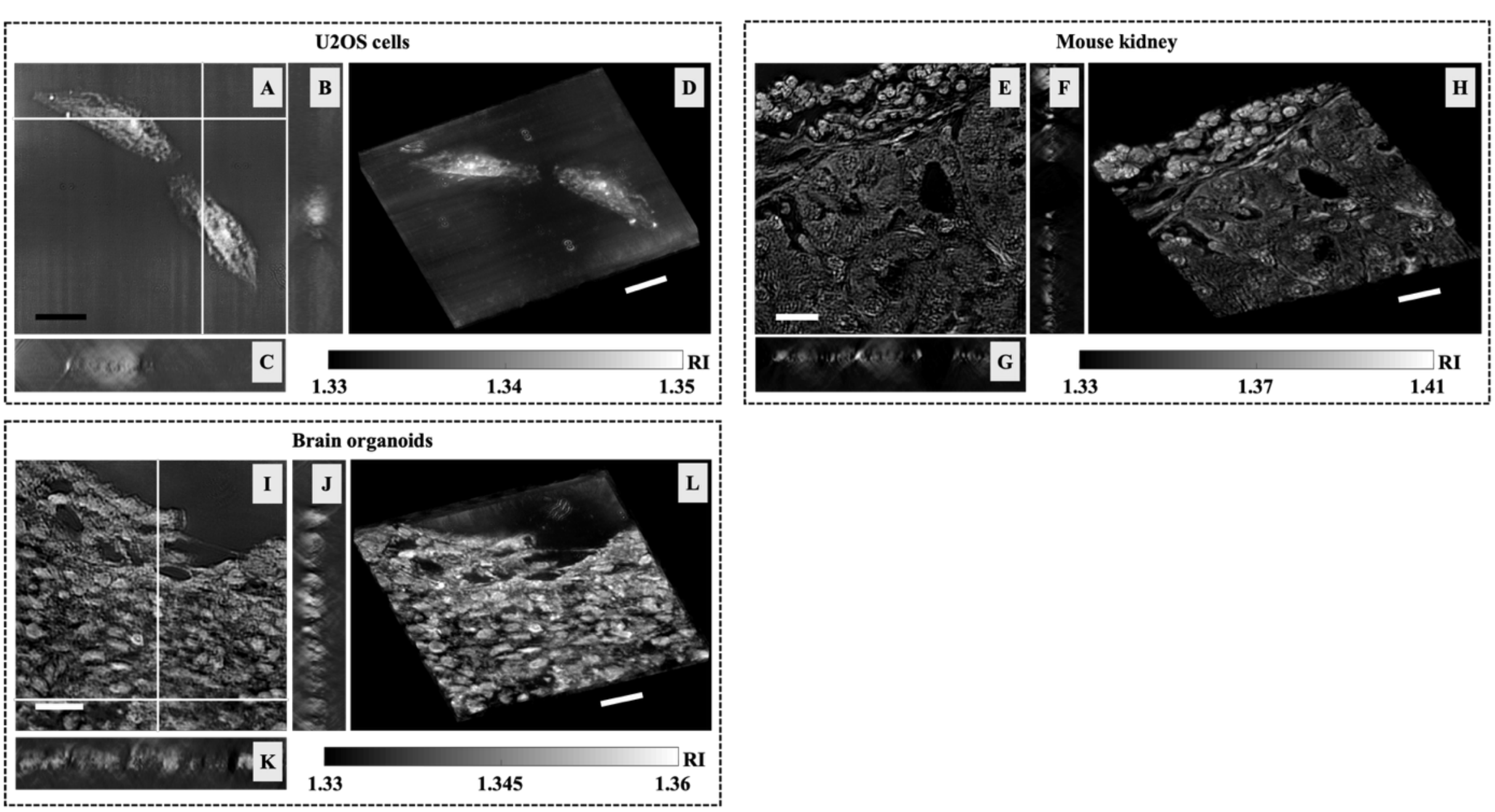

***Figure 4 – Thin biospecimens imaging.*** *U2OS cells: X-Y cross-section of RI map* ***(A)****, X-Z cross-section of RI map* ***(B)****, Y-Z cross-section of RI map* ***(C)****, and RI map in isometric view* ***(D)****. Mouse kidney tissue: X-Y cross-section of RI map* ***(E)****, X-Z cross-section of RI map* ***(F)****, Y-Z cross-section of RI map* ***(G)****, and RI map in isometric view* ***(H)****. Thin brain organoids: X-Y cross-section of RI map* ***(I)****, X-Z cross-section of RI map* ***(J)****, Y-Z cross-section of RI map* ***(K)****, and RI map in isometric view* ***(L)****. All scale bars correspond to 20 micrometers.*

### 2.3. 3D RI imaging of thick section of brain organoid

So far, the results have demonstrated that partial coherence LS-ODT performs effectively on thin biological samples, such as U2OS cells, thin tissue, and thin brain organoid sections. To further evaluate the capability and robustness of the approach, we extended the experiments to a 120 µm-thick tissue section of brain organoids. This allows assessment of the system's performance under more challenging imaging conditions involving greater sample thickness and increased scattering.

As a first step in evaluating the robustness of the system for 3D RI imaging of thick samples, we investigated the variation in interference fringe contrast as a function of depth. This analysis provides insight into the system's ability to maintain fringe visibility and phase stability across different axial planes, which is critical for reliable tomographic reconstruction in highly scattering or layered biological specimens. To investigate fringe visibility as a function of imaging depth, a series of mesh-type interferograms were captured by vertically scanning the sample stage from 0 to 300 µm, in 5 µm steps. Figures 5A–5C show representative interferograms acquired from a thick brain organoid section at depths of 50, 150, and 250 µm, respectively. The fringe contrast was quantified using a Fourier transform-based method [43], and the results were plotted as a function of depth, as shown in Fig. 5D.

The measured interference contrast is found to be approximately equal to 0.45 and the graph clearly demonstrate that the contrast does not deteriorate with increasing imaging depth. The relatively low contrast is primarily attributed to the presence of polarizer P4 (see Fig. 2), which alters the relative intensities of the interfering beams at the camera plane. This depth independent contrast highlights the system’s stability and depth-penetration capability, which is essential for accurate 3D RI reconstruction in thick, scattering biological samples.

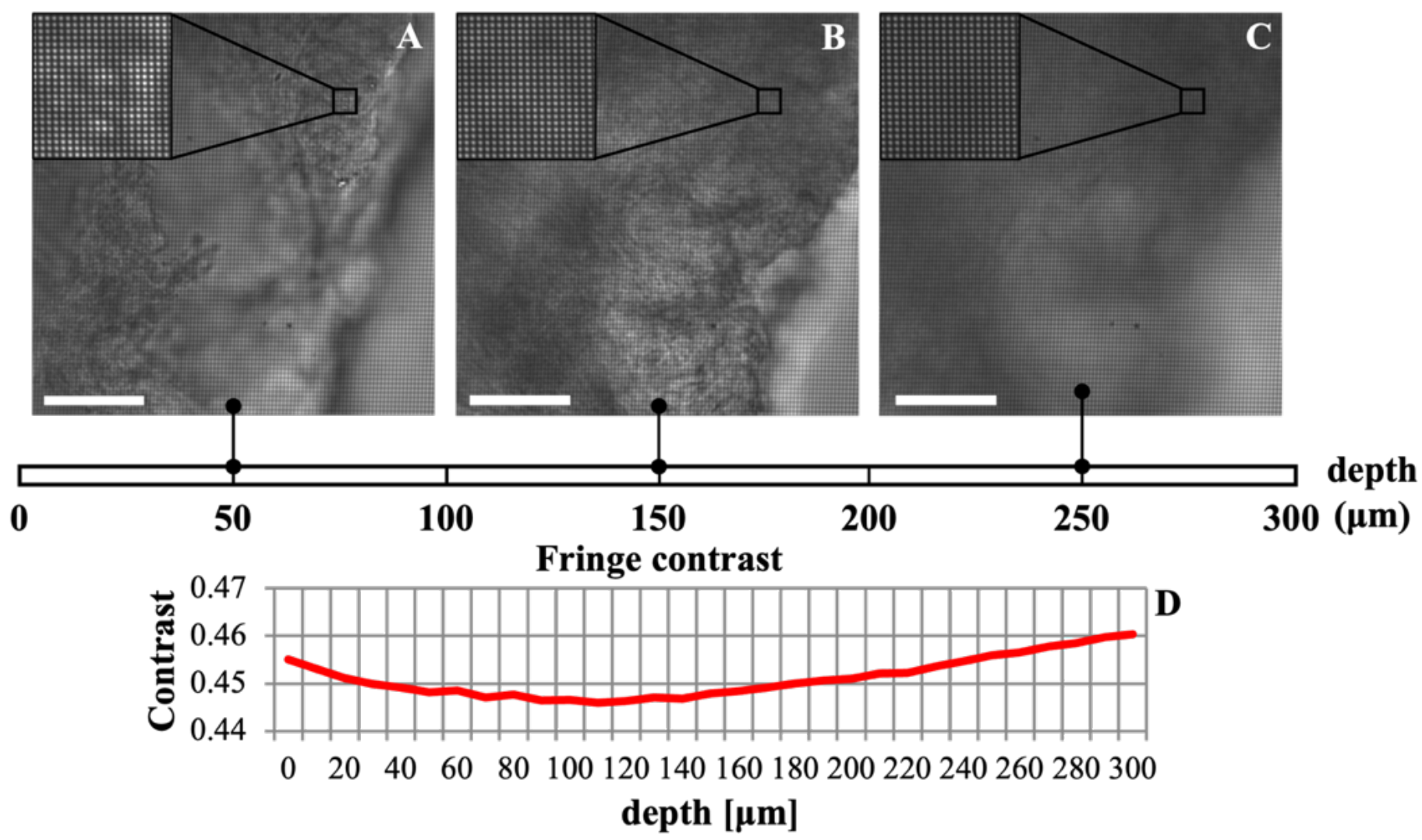

***Figure 5 – Depth vs fringe contrast.*** *Z-scan of thick brain organoid sample, confirmation of high quality of the fringes at every depth. Interferograms from* ***A)*** *10 µm deep,* ***B)*** *60 µm deep,* ***C)*** *110 µm deep with zoomed views. Chart of fringe contrast as a function of depth* ***(D)****. All scale bars correspond to 50 micrometers.*

Next, we performed RI reconstruction on a 120 µm-thick brain organoid tissue section to evaluate the depth-resolving capacity of the LS-ODT system in highly scattering biological media. For this analysis, interferograms were acquired at seven axial (Z-axis) planes, spaced 10 µm apart. At each level, a 10 µm-thick RI slice was reconstructed, encompassing an imaging range of ± 5 µm. These slices were subsequently stacked to form a composite 3D RI volume spanning a total depth of 80 µm, as shown in Figs. 6A – 6H.

The reconstructed volume demonstrates consistent phase contrast and well-defined internal features across all depths, validating the method’s robustness for volumetric imaging in dense biological samples. Importantly, this result confirms that partial coherence LS-ODT preserves interference fringe visibility and maintains RI reconstruction quality even in regions with high cellular packing and optical scattering, conditions typical of maturing brain organoids.

From a biological perspective, the ability to resolve microarchitecture throughout the entire 80 µm thickness offers key insights into neurodevelopmental spatial organization. The RI maps reveal layered structural heterogeneity consistent with radial organization of progenitor zones and differentiated neuronal territories. Denser, RI-elevated regions likely correspond to clusters of post-mitotic neurons and emerging neuropil, whereas more homogeneous zones may indicate proliferative stem/progenitor cell layers. These observations align with expected morphological transitions during cortical development and showcase the utility of label-free RI tomography in preserving and revealing such delicate architectures.

For comparative benchmarking, the same organoid sample was imaged using a commercial Nanolive system, which relies on a non-common-path Mach–Zehnder interferometric geometry. As expected, Nanolive failed to produce consistent or biologically interpretable RI reconstructions beyond ~30 µm depth (Figs. 6I – 6L). This limitation arises from the system's susceptibility to multiple scattering, speckle noise, and temporal phase instability, all of which degrade image quality in thick specimens.

Together, these findings demonstrate that partial coherence LS-ODT enables high-fidelity 3D imaging across extended tissue depths, addressing a major unmet need in organoid research. The system offers a non-destructive alternative to physical sectioning, making it ideally suited for the study of intact brain organoids, where maintaining spatial context and cytoarchitecture is critical. Ongoing studies will further explore the system's maximum effective imaging depth, particularly for specimens exceeding 300 µm in thickness.

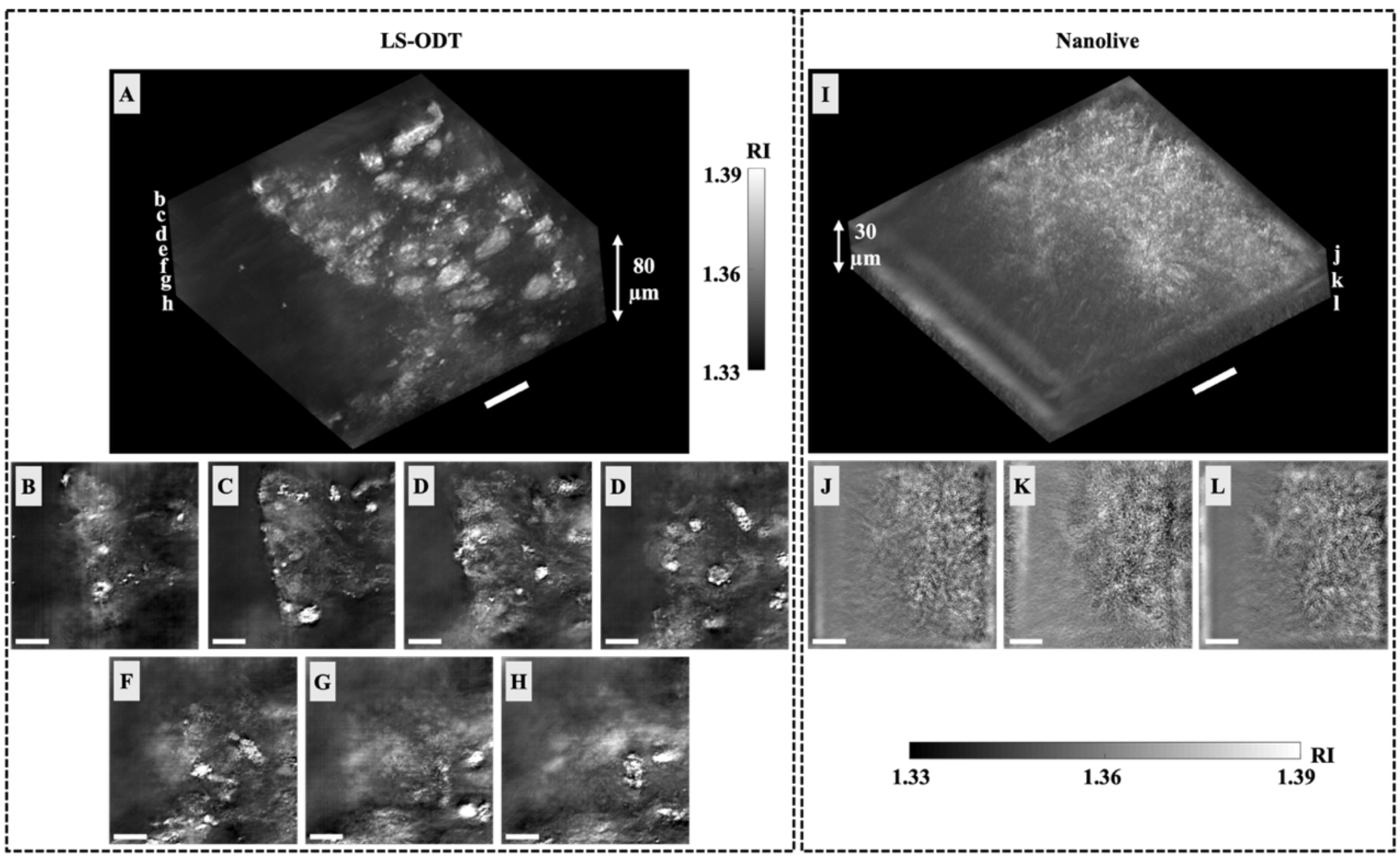

***Figure 6 – Imaging of an 80-micron thick brain organoid tissue.*** *RI map from LS-ODT setup in isometric view **(A)**. Multiple X-Y cross-sections of RI map at following depths: 10 µm **(B)**, 20 µm **(C)**, 30 µm **(D)**, 40 µm **(E)**, 50 µm **(F)**, 60 µm **(G)**, and 70 µm **(H)**. RI map from Nanolive setup in isometric view **(I)**. Multiple X-Y cross-sections of RI map at following depths: 10 µm **(J)**, 20 µm **(K)**, and 30 µm **(L)**. All scale bars correspond to 20 micrometers.*

### 2.4. Spatial Identification of Neurogenic Compartments in Early Brain Organoids:

To further assess the biological relevance of RI imaging in early-stage brain organoids, we performed high-resolution, correlative analysis of selected subregions from a human brain organoid. Figure 7 shows the entire human brain organoid at day 30 of differentiation, labeled with DAPI (nuclei; blue) and MAP2 (mature neurons; red). This early-stage organoid demonstrates partial neuronal maturation, with distinct regional differences in neuronal density. This pattern is consistent with early cortical development, where progenitor zones coexist alongside emerging neuron-rich areas.

The wide-field image was acquired using a 20×/0.7 NA objective, and three regions of interest (ROIs) were selected for high-resolution correlative analysis. ROI 1, marked by a white rectangle, was imaged using DeltaVision (deconvolution) fluorescence microscopy with a 60×/1.42 oil immersion objective, while ROIs 2 and 3 (yellow and green rectangles, respectively) were further imaged using the LS-ODT system. These ROIs were chosen to sample distinct developmental regions: ROI 2 corresponded to an area with a higher concentration of MAP2$^+$ mature neurons, whereas ROI 3 targeted an area with low MAP2 signal, indicative of undifferentiated or progenitor cell populations. Notably, edge regions of the organoid, which showed high MAP2 labeling, were excluded from detailed imaging to focus on subtler internal patterning.

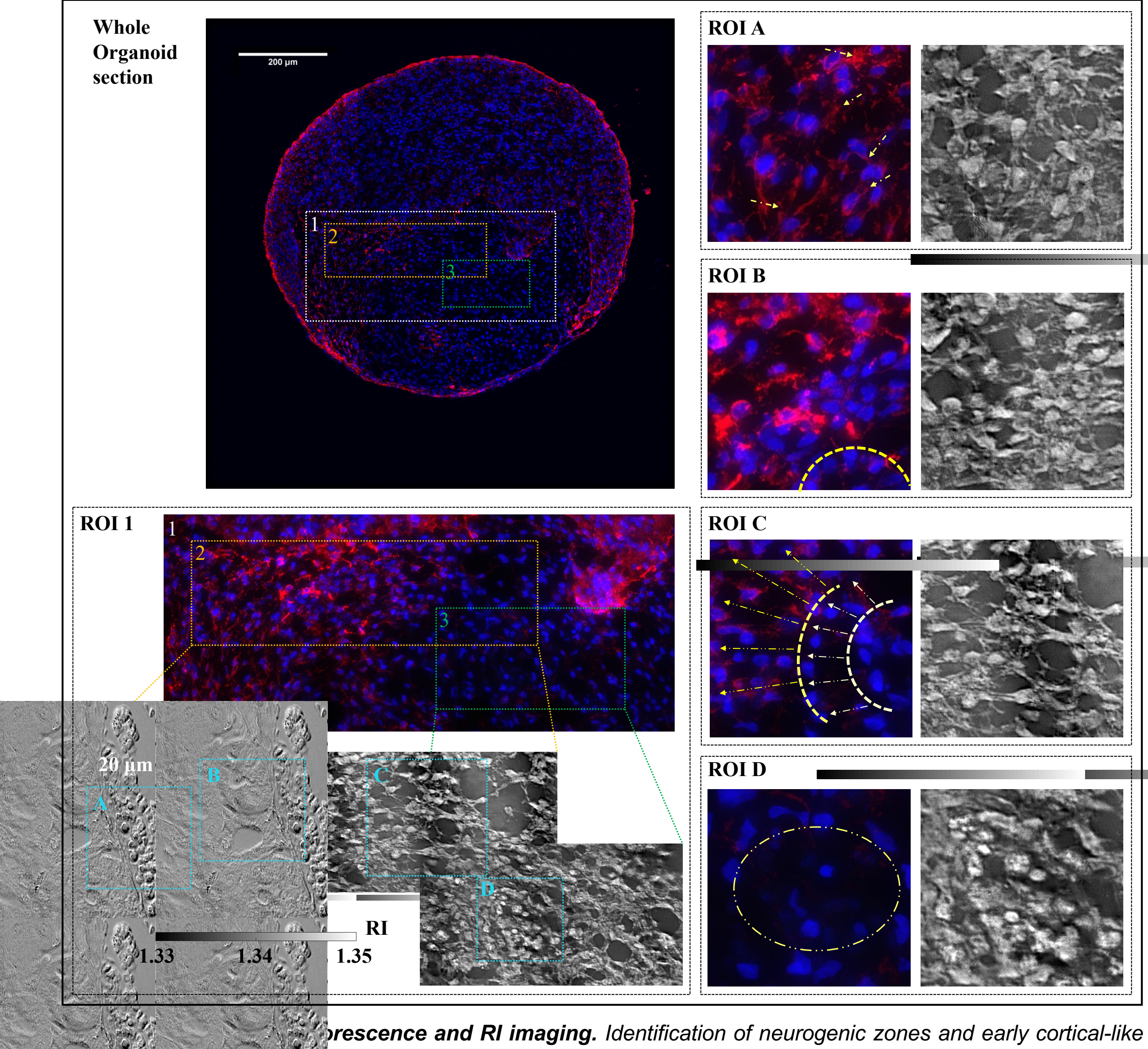

***…orescence and RI imaging.*** *Identification of neurogenic zones and early cortical-like …ids using correlative fluorescence and RI imaging. The fluorescent image of the entire …e top left corner. ROI 1, ROI 2 and ROI 3 are marked with white, yellow and green …Is A-D are marked with cyan rectangles on RI map of ROI 2 and ROI 3. ROI A shows mature neurons (MAP2$^+$, red) with long, extended processes forming early neuronal networks (arrowheads), suggesting developing synaptic connectivity. ROI B highlights a subventricular zone (SVZ)-like neurogenic niche (dashed yellow line), characterized by dense nuclear labeling (DAPI$^+$, blue) and absence of MAP2 signal, indicative of a proliferative progenitor region. ROI C demonstrates outward migration of differentiated neurons from the SVZ, mimicking early cortical layering. Arc-like dashed lines mark emerging laminar zones, while dashed arrows indicate* *the radial trajectory of migrating neurons. ROI D shows a circular neural tube-like structure (dashed circle), lined with neural stem cells (DAPI$^{+}$/MAP2$^{-}$), forming a rosette-like pattern consistent with neuroepithelial organization. All fluorescent images were acquired on commercial DeltaVision fluorescence microscope.*

Figure 7 presents correlative imaging of ROI 1 (fluorescence) with RI maps from ROI 2 and ROI 3, respectively. Across both regions, the measured refractive index (RI) ranged from 1.33 to 1.35, reflecting a relatively narrow distribution — consistent with the compact and still-developing tissue architecture typical of early-stage organoids. Interestingly, although ROI 3 contained fewer MAP2$^{+}$ cells, its RI values remained comparable to ROI 2, suggesting a dense cellular environment composed primarily of neural progenitors. This aligns with the known biology of early organoids, where radial glial and intermediate progenitor cells form dense layers prior to neuronal migration and laminar organization.

To further assess the biological relevance of RI imaging in early-stage brain organoids, we performed additional correlative analysis of small subregions (Fig. 7, ROIs A-D) within ROI 2 and ROI 3. Four distinct ROIs were selected based on combined RI and fluorescence features to capture hallmarks of neurodevelopmental organization. ROI A revealed a neuron-enriched region displaying mature MAP2$^{+}$ cells with extended processes, indicative of early neuronal network formation. In contrast, ROI B delineated a densely packed zone resembling a subventricular zone (SVZ)-like progenitor niche, with high nuclear density and minimal MAP2 signal. ROI C exhibited features of radial neuronal migration, with MAP2$^{+}$ differentiated cells streaming outward in a layered, arc-like pattern reminiscent of early cortical lamination. Finally, ROI D highlighted a distinct circular zone lined by DAPI$^{+}$/MAP2$^{-}$ cells, suggestive of a neural tube-like structure, potentially representing a neuroepithelial rosette or proliferative stem cell ring. These features, together, recapitulate fundamental aspects of early human brain development and demonstrate the ability of LS-ODT, when combined with fluorescence microscopy, to reveal functionally distinct neurogenic compartments within organoids based on intrinsic optical properties and spatial organization. Small extension of correlative ROI 3 analysis is provided in Supplementary Figure S8.

## 3. Discussion and conclusion:

In this study, we proposed an interferometric-based common path ODT system with reduced spatial coherence as a novel solution for quantitative and label-free imaging of thick, strongly scattering biological samples. The system, known as LS-ODT, effectively suppresses the impact of multiple scattering on imaging quality and fringe contrast. Common path architecture provides high temporal stability, while reduction of spatial coherence ensures high spatial phase and RI sensitivity through speckle noise reduction and overall beam quality improvement. The combination of these features makes LS-ODT an exquisite alternative to conventional ODT solutions, especially in the field of label-free histology.

Robustness of this approach has been confirmed by a series of experiments on a wide variety of samples. RI reconstruction of artificial cell phantom corroborated positive effect of coherence reduction on RI sensitivity. Furthermore, investigation of an object with known RI distribution enabled us to reliably benchmark the LS-ODT system in terms of the accurate reconstruction of absolute RI values. Experiments on U2OS cells, mouse kidney tissue and brain organoids showed that this approach can serve as a reliable tool for label-free imaging of biological samples regardless of their structure, shape and optical density. LS-ODT offers high quality of the RI reconstruction for large range of sample thicknesses, up to 80 μm. The true thickness limitation of the system has not been explored yet and in future studies, performance of LS-ODT will be investigated on over 300 μm thick specimens. Currently, there are no clear signs of depth-dependent reconstruction quality decrease or fringe visibility deterioration at 80 μm depth.

From a biological standpoint, the ability to perform quantitative, volumetric imaging in intact, label-free brain organoid sections offers a unique opportunity to examine spatial organization within developing neural tissues. In our experiments, thin sections of brain organoids revealed relatively narrow RI variation (~0.03), likely reflecting the uniform cellular density and immature neuroepithelial state of early organoid development. These observations align with the predominance of undifferentiated or partially differentiated neural progenitors, which lack the heterogeneous extracellular matrix and layered structure seen in mature cortical tissue.

More notably, LS-ODT enabled 3D imaging of 80 μm-thick brain organoid sections, resolving depth-dependent structural heterogeneity across the volume. These thick section reconstructions revealed subtle layering and RI gradients that likely correspond to zones of neural differentiation, with denser regions potentially indicating post-mitotic neurons and neuropil, and more homogeneous areas representing proliferative progenitor domains. Such internal patterning is critical to understanding the temporal and spatial aspects of neurodevelopment, especially in models of cortical organization, radial glia migration, and early brain patterning.

Correlative imaging using fluorescence markers (DAPI, MAP2, GFAP) further validated these findings. Regions with higher MAP2 expression, indicating mature neurons, displayed locally elevated RI values, while $MAP2^{-}$/$DAPI^{+}$ regions associated with stem cell-rich zones showed more uniform RI textures. Importantly, the ability to map these molecularly defined regions onto intrinsic RI contrast enables researchers to identify developmental compartments without relying exclusively on immunostaining or physical sectioning.

Thus, the label-free nature of LS-ODT, coupled with its compatibility with fluorescence microscopy, makes it especially valuable for organoid research, where structural preservation and multiplex readouts are essential. As brain organoids evolve to model not just early development but also conditions like neurodegeneration, inflammation, and immune system interactions, there is a growing need for non-destructive 3D imaging methods. These tools are crucial for advancing our understanding in both basic neuroscience and translational research.

The LS-ODT system, despite its advantages, still faces a few challenges, among which most of them will be targeted during further development. The proposed approach requires a rather voluminous optical system in comparison to grating-based or intensity based ODT solutions. The RI reconstruction is not artifact-free and ringing effect can be observed especially at abrupt sample edges. The combination of LS-ODT experimental system and more advanced 3D reconstruction algorithms (including Full-Wave Inversion and modified Born-series) can bring potential further increase in RI reconstruction quality. Even though the effect of multiple scattering has been largely reduced compared to conventional ODT techniques, it is not clear to what extent, and it can be further quantified in a separate study. A long acquisition time and lack of incubator makes the current implementation of the system unsuitable for any live sample imaging. Use of galvo mirrors can decrease the acquisition time and will allow for observation of dynamic living samples.

## 4. Materials and methods:

### 4.1. Experimental setup

Most of the experimental system's description has been included in the introduction section and in Fig. 2; therefore, this subsection will be mostly dedicated to pure technical details of the LS-ODT setup. All images were acquired with 60×/1.2NA Olympus objective lens with water immersion. As a light source we have utilized 700mW diode laser from OptLasers company. The wavelength of the light was equal to 638 nm. All images were captured with Hamamatsu Orca Flash 4.0 CMOS camera. Matrix of the camera consists of 2048×2048 pixels, 6.5×6.5 μm each. Acquisition of entire 360-degree set of interferograms was performed via Thorlabs Kinesis rotation stage (part # PRM1/MZ8) and took 5 minutes and 14 seconds. Acquisition time is not a limitation of the system itself and can be improved with the application of more advanced scanning devices like galvo mirrors, DMDs, and MEMS scanners. The overall magnification of the system is 65×, corresponding to a pixel size of 100 nm at the sample plane. Theoretical maximal FoV of the system is therefore equal to 204.8×204.8 μm. However, removal of the $4^{th}$ beam with a polarizer caused geometric issues for certain set of the angles, thus limiting the effective FoV to around 120×120 μm. This is not an inherent issue of the system itself, and it will be addressed in further development. The shear between the beams is equal to 328 nm in X-axis and 349 nm in Y-axis. Details regarding calculation of the shear are given in supplementary section and in supplementary Figure 3.

### 4.2. Role of coherence in ODT:

The coherence properties of the light source play a crucial role in quantitative phase microscopy (QPM) [44-48] and ODT [49, 50]. Incoherent light sources, such as halogen lamps and LEDs, improve interference quality by mitigating speckle noise and parasitic fringes, leading to increased spatial phase and RI sensitivity. However, its shorter temporal coherence lengths make it significantly more challenging to generate stable interference fringes. Additionally, incoherent light sources are generally unable to produce high-density interference fringes across the entire field of view (FOV) [46], except in grating-based QPM systems [51]. This limitation necessitates the acquisition of multiple phase-shifted fringes for accurate, lossless phase recovery, ultimately reducing the imaging speed. To address this challenge, DSI is used that generates high-density interference fringes across the entire camera FOV while maintaining improved interference fringe quality comparable to that of the incoherent light sources [46]. This approach effectively balances coherence and interference quality, enhancing the overall performance of QPM and ODT systems.

### 4.3. Quantitative phase map reconstruction procedure

The entire data reconstruction procedure is presented on a flowchart in Fig. 8. We have applied Fourier Transform (FT) as the phase reconstruction technique. FT algorithm is chosen due to its most reliable and well-established single frame solution. Multi frame solutions like phase shifting were not considered because they further increase the acquisition time and introduce another degree of freedom to the setup. For each sample, one additional set of background interferograms was acquired (from all angles of illumination). FT method allows for clean and elegant separation of X-axis and Y-axis information via choice of corresponding Fourier peak. For each angle, a total of four differential phase maps are reconstructed, two of them corresponding to X-axis (sample and background) and to Y-axis (also sample and background). After unwrapping stage, background phase map is subtracted from the sample phase map. Subsequently, two-dimensional integration is performed using differential phase maps from X and Y-axis in order to retrieve the true phase map of the sample. The shear values are used as spacing values (dx and dy) in the phase integration step. During the stage of phase reconstruction, amplitude of the interferogram is acquired as well. Integrated phase maps and amplitudes from all angles of illumination (180 angles in total with 2-degree interval) are put together and form a sinogram, which serves as the input for RI reconstruction.

### 4.4. 3D RI map reconstruction procedure

The tomographic reconstruction is performed using the direct inversion (DI) method, also known as direct interpolation, [52] based on first order Rytov approximation [53]. The first step of the algorithm involves computing Rytov-transformed object fields: $u_R = ln(u/u_i)$, where $u_i$ represents the illumination beam generating the corresponding object wave $u$. Next, using Wolf's generalized projection theorem [54], the Rytov field spectra are mapped onto Ewald spheres in the 3D Fourier domain, enabling the reconstruction of the object scattering potential spectrum, $\tilde{O}\,(f_x, f_y, f_z)$.
Due to the limited range of illumination directions and the limited numerical aperture, substantial regions of $\tilde{O}$ remain empty. To address this, the next step applies an iterative Gerchberg-Papoulis (GP) algorithm with a non-negativity constraint [55]. This algorithm iteratively switches between the spatial and Fourier domains, applying different constraints in each. In the spatial domain, the constraint ensures that the refractive index (RI) values remain above (or below) the surrounding medium's RI, $n_0$. In the Fourier domain, the constraint enforces that only the missing regions of $\tilde{O}$ are modified, while the known spectrum remains unchanged. In our case, we run 10 iterations of the GP algorithm.
Finally, the refractive index distribution is evaluated using the formula:

$$n(x,y,z) = n_0\sqrt{1 - \tilde{O}\,(x,y,z)}$$

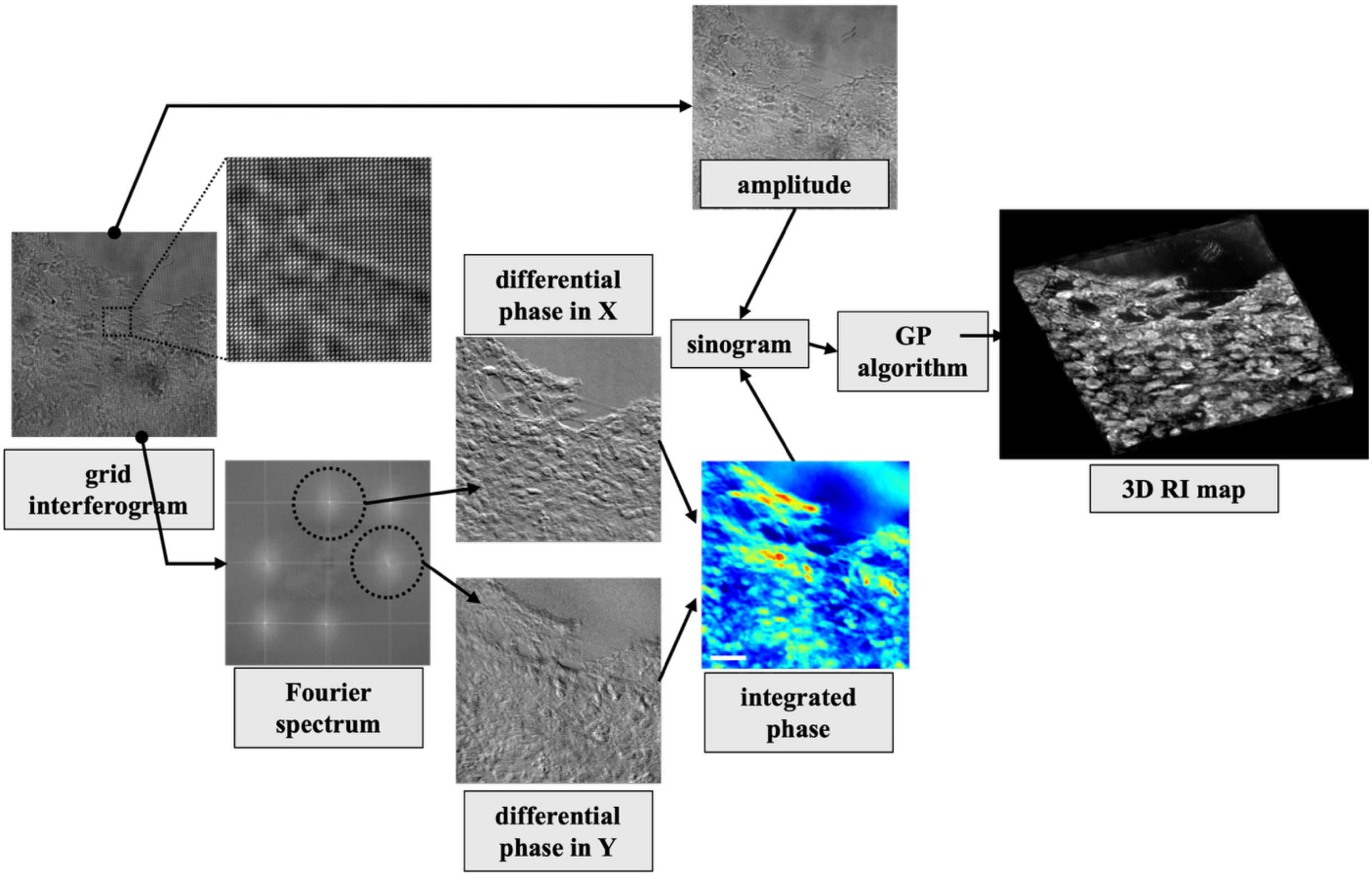

***Figure 8 – Step-by-step flowchart of the RI reconstruction procedure.*** *The amplitude and differential phase maps are retrieved from the raw interferogram via filtration in Fourier domain. Differential phase maps in X and Y axis undergo 2-dimensional integration. Integrated phase maps and amplitudes from 180 angles form a sinogram. 3D RI map is retrieved from the sinogram via Gerchberg Papoulis algorithm.*

### 4.5. Sample preparation:

For the ground truth investigation of the system, we used an artificial cell phantom. The phantom was prepared via three dimensional two-photon photolithography (alternatively known as direct laser writing). In this technique, two-photon absorption leads to solidification of a liquid polymer in a confined volume around the focus of a femtosecond-pulsed laser beam. Scanning the focal region along a given trajectory forms custom, phase-only 3D structures. Some of the structures are designed so that they mimic optical and structural properties of mammalian cells, while the rest serve as resolution tests in X, Y and Z-axis. The RI of the structures range from 1.49 to 1.52. The phantom was printed on a flat glass slide. PDMS spacer was stuck to that surface and covered with a cover glass. The space between the glass substrate and the cover slip was filled with immersion oil of RI equal to 1.46. More details about the design and manufacturing of the phantom can be found in a work from Ziemczonok et.al. [39].

U2OS cells were cultured in a standard DMEM culture media supplemented with 10% FBS and 1% penicillin-streptomycin) and maintained at 37°C in a 5% $CO_2$ incubator. For imaging, cells were seeded onto 1.5 glass bottom Petri dishes and kept overnight in the incubator to adhere on Perti dish surface. Prior to fixation, cells were washed gently with phosphate-buffered saline (PBS) to remove debris and excess media. A 4% paraformaldehyde (PFA) solution was used to fix the sample for 10–15 minutes at room temperature, followed by 2× PBS wash to remove residual fixative.

Kidney from a young mouse was harvested and fixed immediately after collection in 10% neutral-buffered formalin (NBF) for 24–48 hours at room temperature to preserve cellular morphology. After fixation, the tissue was dehydrated through a graded series of ethanol, cleared in xylene, and embedded in paraffin wax. Thin sections of 10 micron were sliced with a microtome and mounted on glass slides.

The sections were then deparaffinized in xylene, rehydrated through decreasing concentrations of ethanol, and stained with hematoxylin and eosin (H&E) for standard histological examination. The same tissue sample slides were examined under LS-ODT microscope to extract 3D RI map of the specimen.

**4.6. Brain Organoid Preparation**

Brain organoids were generated from human foreskin fibroblast (ATCC, SCRC-1041) derived iPSCs, following previously established protocols with minor modifications [56, 57]. Briefly, iPSCs were cultured under feeder-free conditions and dissociated into a single-cell suspension. Cells were seeded into AggreWell 800 plates (STEMCELL Technologies) to promote uniform 3D aggregate formation in neural induction medium (NIM) containing dual SMAD inhibitors and 10 µM ROCK inhibitor (MedChemExpress). After 48 hours (day 2), the medium was replaced with fresh NIM lacking ROCK inhibitor and was renewed daily until day 6.

On day 7, the developing neuroepithelial aggregates were embedded in basement membrane extract (Geltrex, Life Technologies), pipetted as dome structures (~10–15 µL) in 100 mm Petri dishes. The domes were allowed to gel for 20 minutes at 37°C before adding neural expansion medium (NEM). Organoids were maintained in NEM with daily media changes until day 10.

From day 11, the domes were carefully detached and transferred to 12-well ultra-low attachment plates (Corning) and cultured in suspension. Medium was replaced every other day until day 24. On day 25, the cultures were transitioned to neural differentiation medium (NDM) to promote cortical maturation. Media changes were carried out every 3–4 days, and the organoids were maintained in NDM until day 30, at which point they were considered relatively developing structures and suitable for imaging. Organoids were fixed with 4% paraformaldehyde (PFA) overnight at 4°C, washed with PBS and cryoprotected with 30% sucrose overnight. Depending on the imaging experiment, organoids were either: Thin sectioned (5–10 µm) or as thick sections (~120 µm), using a cryostat, for volumetric refractive index tomography. For correlative imaging, organoid sections were immunostained with DAPI (nuclear label), MAP2 (mature neuron marker), and GFAP (astrocyte marker) and imaged using DeltaVision deconvolution microscope.

**Data availability statements:**

The datasets generated and/or analyzed during the current study are available from the corresponding author upon reasonable request.

**Funding:**

A.A. acknowledges FRIPRO Young (project # 345136) funding from Research Council of Norway. B.S.A. and P.G. acknowledges the funding from UiT Thematic Funding (NASAR). M.V. acknowledges the funding from VINNOVA Sweden 2023-01471. UN, MV and BSA acknowledge funding received from the European Union's HORIZON Research and Innovation Actions under grant agreement No. 101191315. B.S.A. and A.A. acknowledges support from the UiT Talent Innovation Grant, Research Council of Norway (Verification Project: 360778). Views and opinions expressed are, however, those of the authors only and do not necessarily reflect those of the European Union. The European Union cannot be held responsible for them. UN also acknowledge funding received from The Swedish Research Council grants 2021-01756 and Karolinska Insititute Consolidator Grants (2-117/2023). M.T. acknowledges project # 2023/48/Q/ST7/00172 funding from Narodowe Centrum Nauki. The publication charges for this article have been funded by a grant from the publication fund of UiT The Arctic University of Norway.

**Author Contribution:**

A.A. conceptualized the idea and designed the optical configuration. A.A. and B.S.A supervised and conceived the project. P.G and A.A. developed the experimental setup. P.G. performed the experiments, analyzed the data, and prepared the figures. P.G. and A.A. mainly wrote the manuscript. J.W. wrote the reconstruction software and helped in the tomographic reconstruction. V.D., M.V., and U.N. prepared the biological samples and provided biological insights. M.V. contributed to the development and insights into the brain organoid work and P.Z. and M.T. contributed to the insights into the ODT. B.S.A. secured the funding for the experimental work. All authors reviewed and edited the manuscript.

**Conflict of interest:** Azeem Ahmad and Balpreet Singh Ahluwalia has submitted a patent application to protect the invention of LS-ODT (Patent Application No. 2513325.7). All other authors declare no competing interests.

**Ethics statement:** All procedures involving human iPSCs were conducted in accordance with applicable ethical standards and regulations. Ethical approval was not required for this study according to the decision of the Swedish Ethical Review Authority (Etikprövningsmyndigheten, Uppsala, Sweden; Dnr 2024-05169-01).

# Supplementary Information

## Lateral shearing optical diffraction tomography of brain organoid with reduced spatial coherence

Paweł Gocłowski[1], Julianna Winnik[2], Vishesh Dubey[1], Piotr Zdańkowski[2], Maciej Trusiak[2], Ujjwal Neogi[3], Mukesh Varshney[3], Balpreet S. Ahluwalia[1,4 #], Azeem Ahmad[1,*, #]

*[1]Department of Physics and Technology, UiT The Arctic University of Norway, 9037 Tromsø, Norway*
*[2]Warsaw University of Technology, Institute of Micromechanics and Photonics, 8 Sw. A. Boboli St., 02-525 Warsaw, Poland*
*[3]The Systems Virology Lab, Division of Clinical Microbiology, Department of Laboratory Medicine, Karolinska Institutet, Stockholm, Sweden.*
*[4]Department of Clinical Science, Intervention and Technology, Karolinska Institute, Stockholm, Sweden*

**Corresponding author: ahmadazeem870@gmail.com, balpreet.singh.ahluwalia@uit.no*
*Email: pawel.goclowski@gmail.com*

*[#]Shared authors*

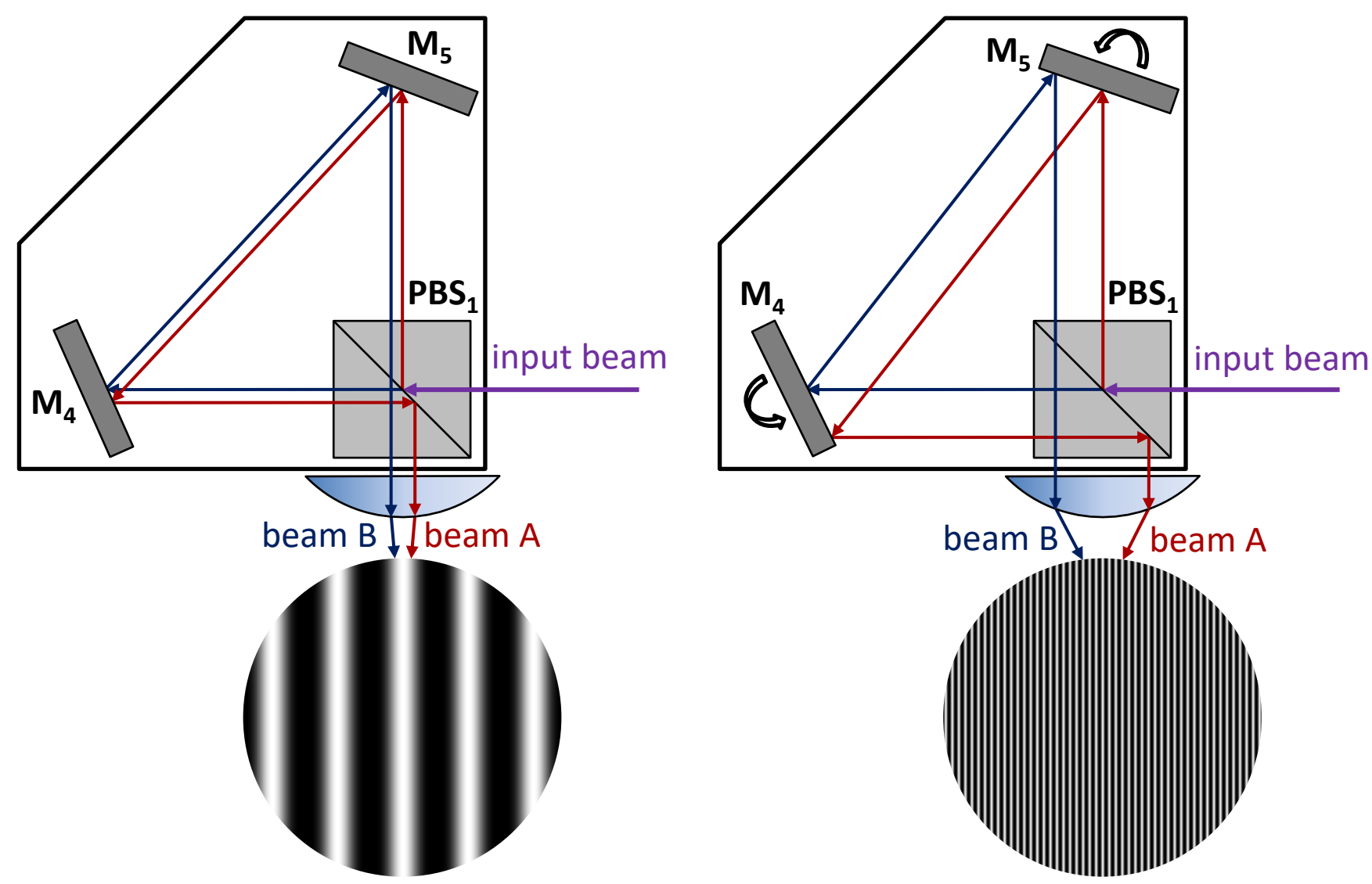

***Supplementary Figure S1 –*** *Working principle of the Sagnac interferometer. The input beam (purple) is splitted by polarizing beamsplitter $PBS_1$ into two beams with equal intensities. The beams get reflected from mirrors $M_4$ and $M_5$ and recombine at $PBS_1$. The rotation of both mirrors in the same direction (clockwise or counterclockwise) changes the separation between beam A and B – as a result, angle between the beams after Sagnac unit increases and so does the fringe density.*

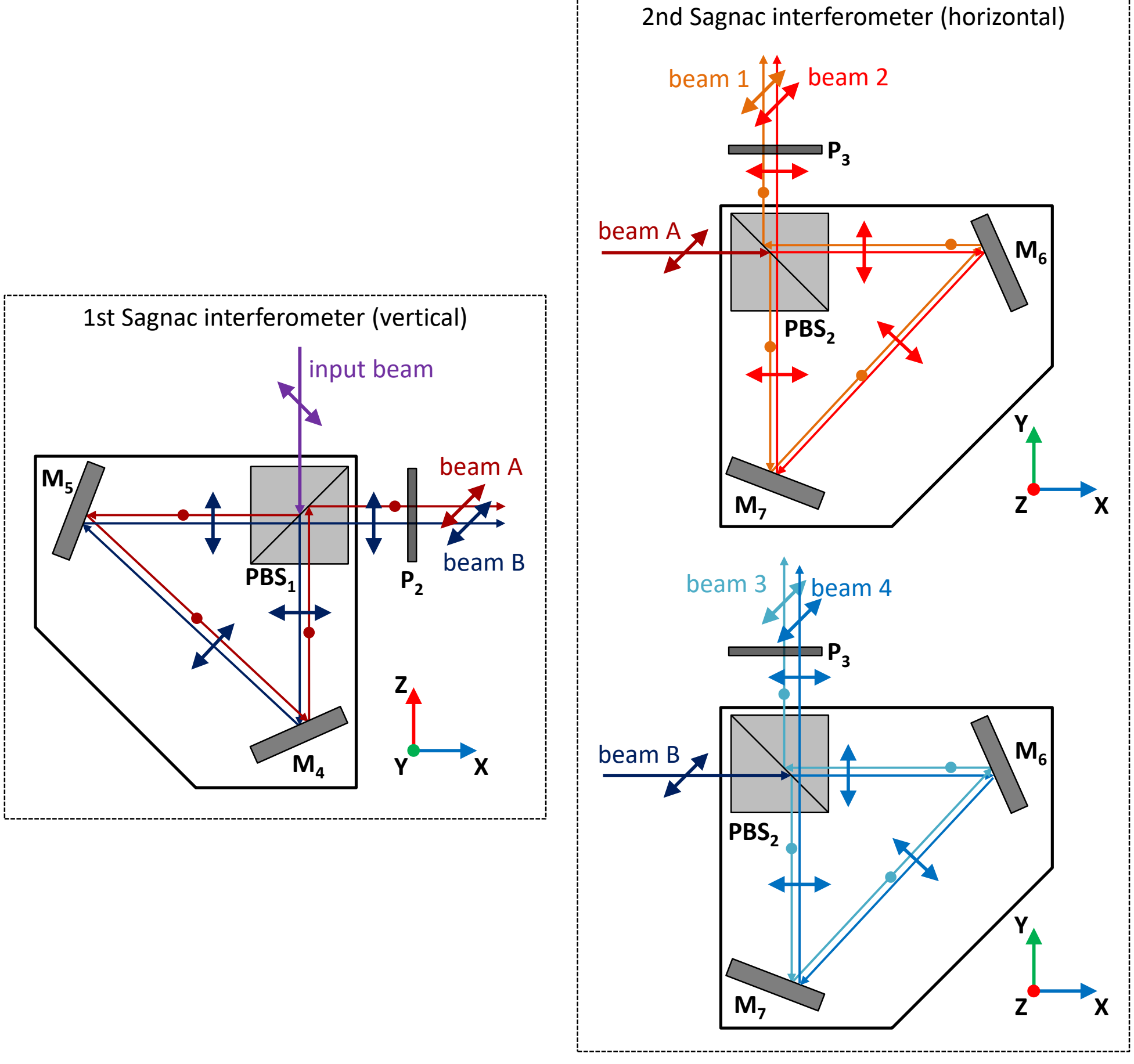

***Supplementary Figure S2*** *– Ray diagram of LS-ODT unit including polarization states. Input beam (purple) enters the $1^{st}$ Sagnac interferometer and gets split at $PBS_1$ into two beams with orthogonal polarizations - beams A (dark red) and B (dark blue). After being reflected from mirrors $M_4$ and $M_5$, beams get recombined at $PBS_1$ and their polarization state is unified by linear polarizer $P_2$. Further propagation of beam A and B through the $2^{nd}$ Sagnac interferometer is shown separately to increase clarity of the scheme. Beam A gets split at $PBS_2$ into two beams with orthogonal polarizations - beams 1 (orange) and 2 (red). Beam B gets split at $PBS_2$ into two beams with orthogonal polarizations - beams 3 (light blue) and 4 (blue). After being reflected from mirrors $M_6$ and $M_7$, beams 1, 2, 3 and 4 get recombined at $PBS_2$ and their polarization state is unified by linear polarizer $P_3$.*

## Shear investigation:

Shear between the beams has been quantified based on 7 µm polystyrene bead imaging. The RI of the bead was 1.59, while RI of the immersion oil was 1.51. The RI difference was therefore equal to 0.08, which results in 0.56 µm optical path difference on the 7 µm bead. The wavelength of the light was 638 nm (0.638 µm) - it means that 0.638 µm corresponds to $2\pi$ radians and 0.56 µm corresponds to $1.75\pi = 5.51$ radians. With this knowledge it is clear that after the phase integration process, the maximal phase of the bead should be equal to 5.51 radians (Fig. S3 (B, E)). 2D integration uses shear values in X and Y-axis as input parameters dx and dy, and if the phase value is correct after the integration, it means that the shear values are set correctly. We decided to iteratively perform integration with increasing shear values in range from 0 to 1000 nm in order to check which shear corresponds to the most accurate phase value of the bead (5.51 radians). The result of this study is shown in Fig. S3 (C, F); the shear is equal to 328 nm in X-axis and 349 nm in Y-axis.

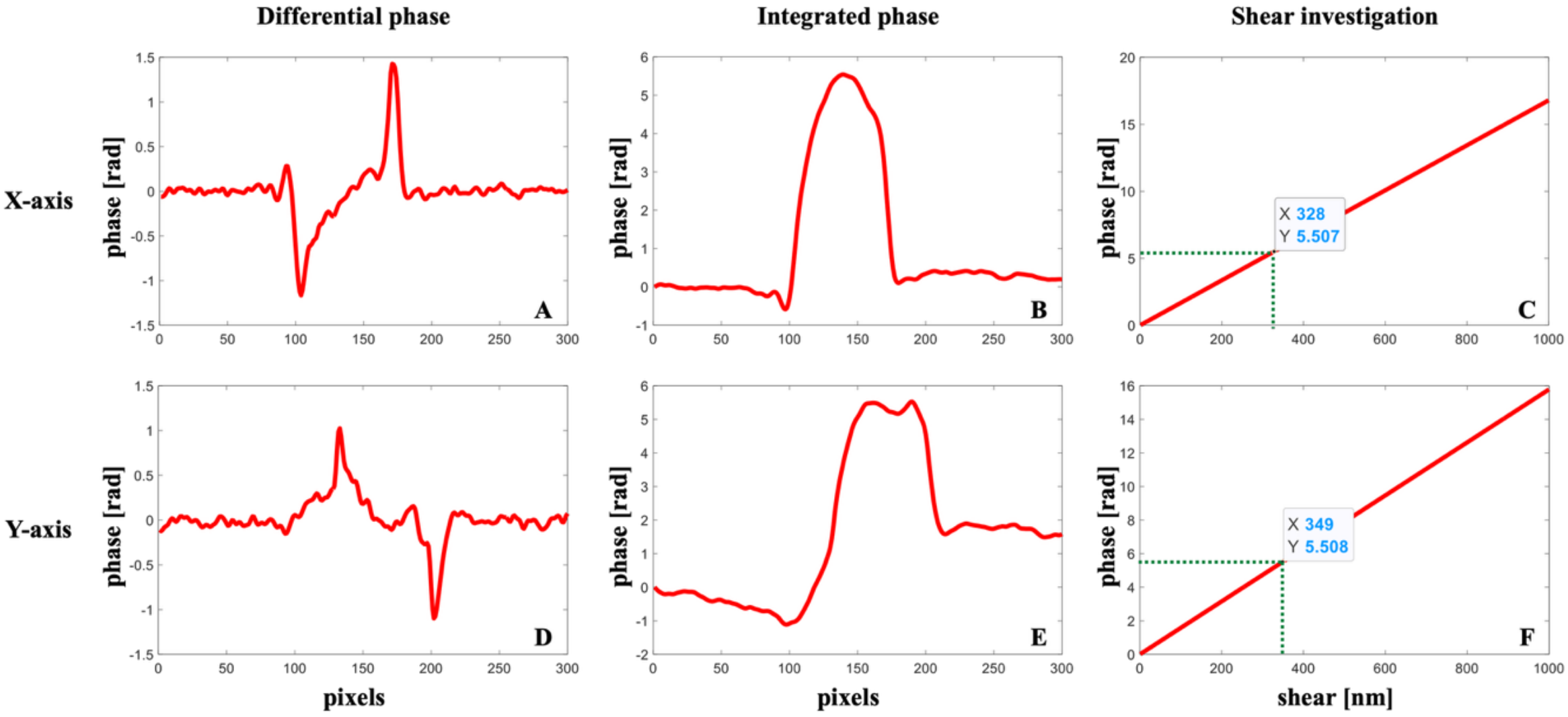

***Supplementary Figure S3 –*** *Shear investigation on 7-micron beads. Line profile of differential phase of the bead in X-axis (A) and Y-axis (D). Line profile of integrated phase of the bead in X-axis (B) and Y-axis (E). Iterative calculation of the shear in X-axis (C) and Y-axis (F), verification for which shear the phase of the bead is correct.*

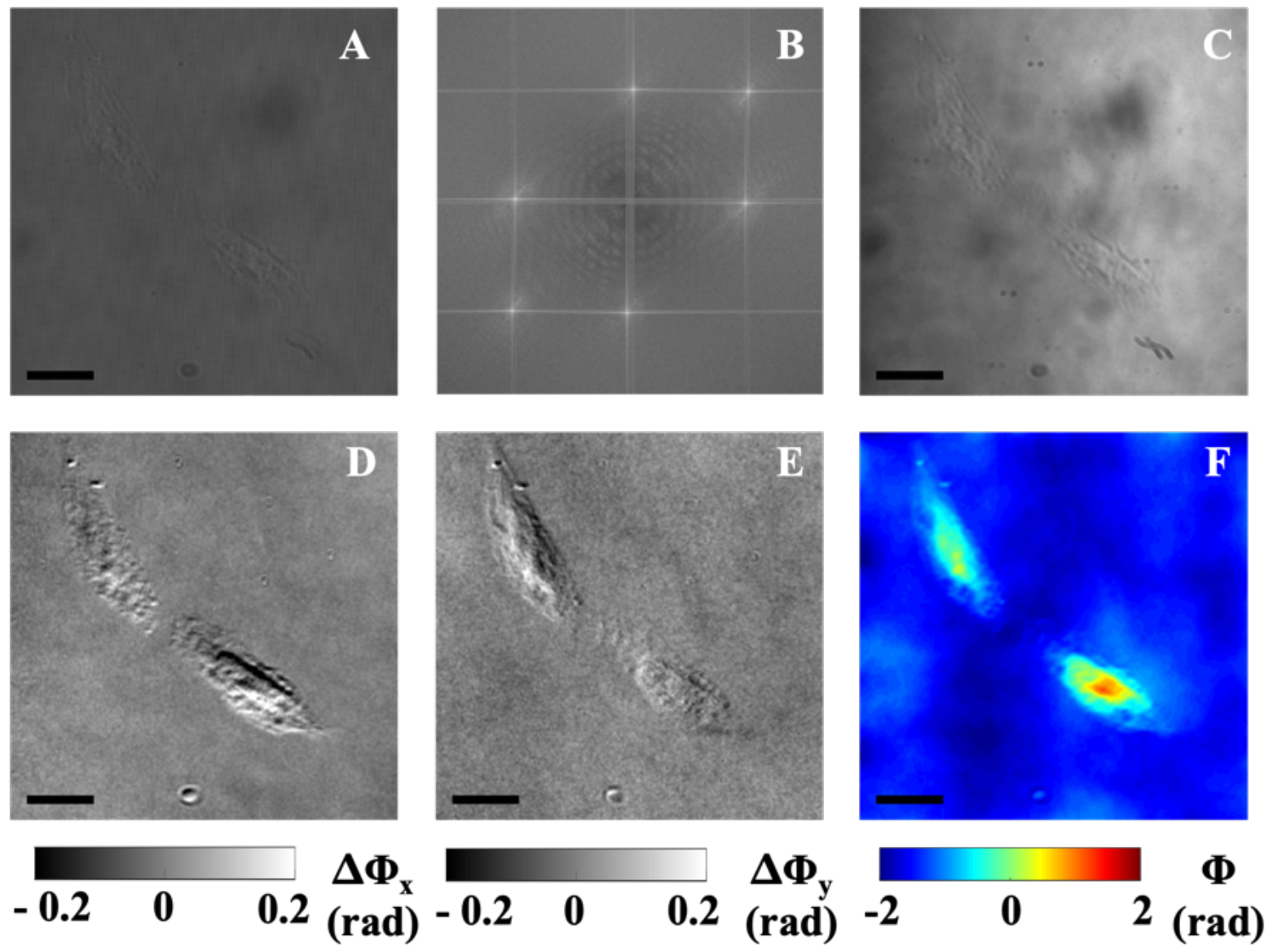

***Supplementary Figure S4 –*** *U2OS cells imaging. Raw interferogram (A), Fourier spectrum (B), amplitude (C), gradient phase in X-axis (D), gradient phase in Y-axis (E), integrated phase (F). All scale bars correspond to 20 micrometers.*

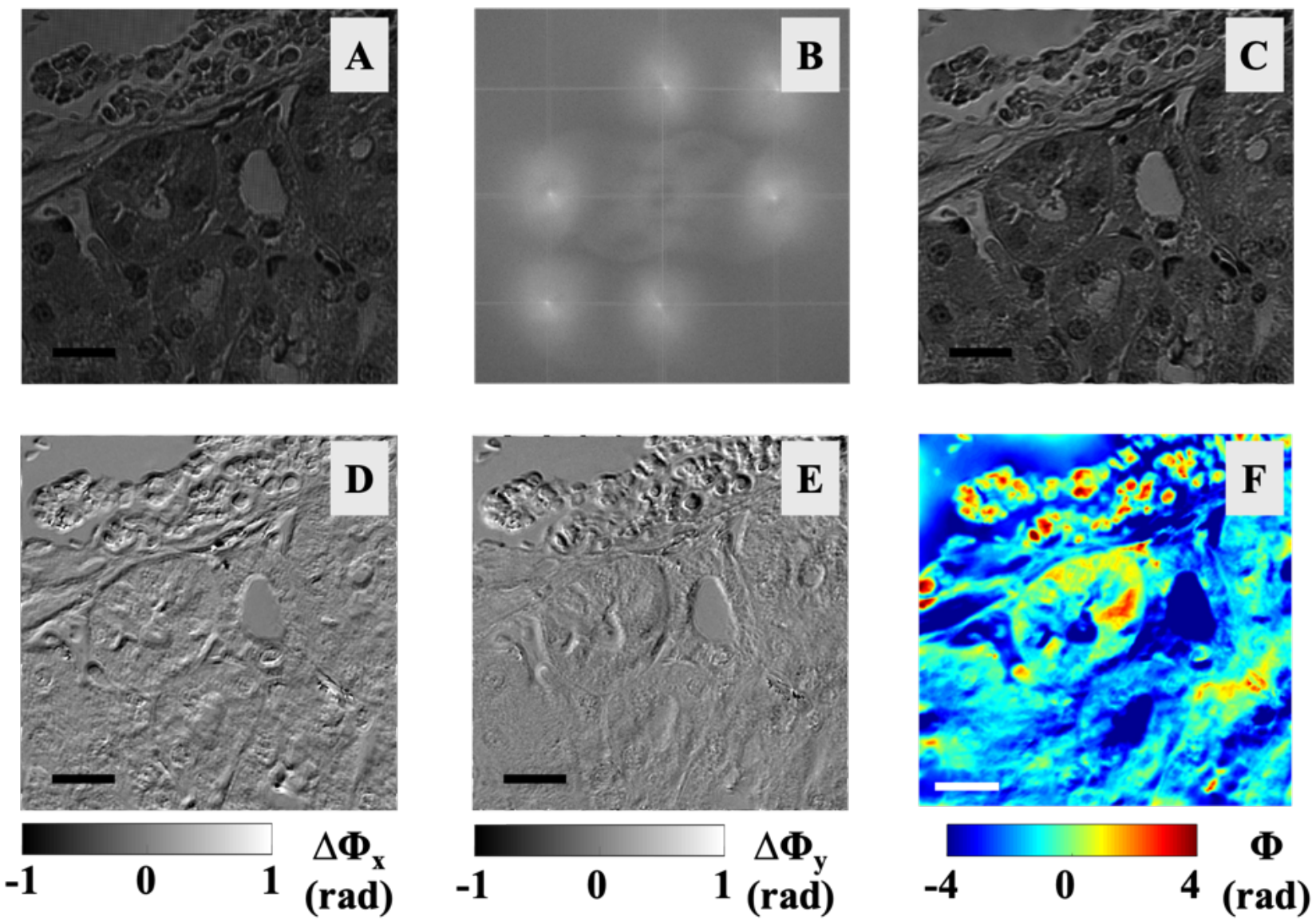

***Supplementary Figure S5 –*** *Mouse kidney tissue imaging. Raw interferogram (A), Fourier spectrum (B), amplitude (C), gradient phase in X-axis (D), gradient phase in Y-axis (E), integrated phase (E). All scale bars correspond to 20 micrometers.*

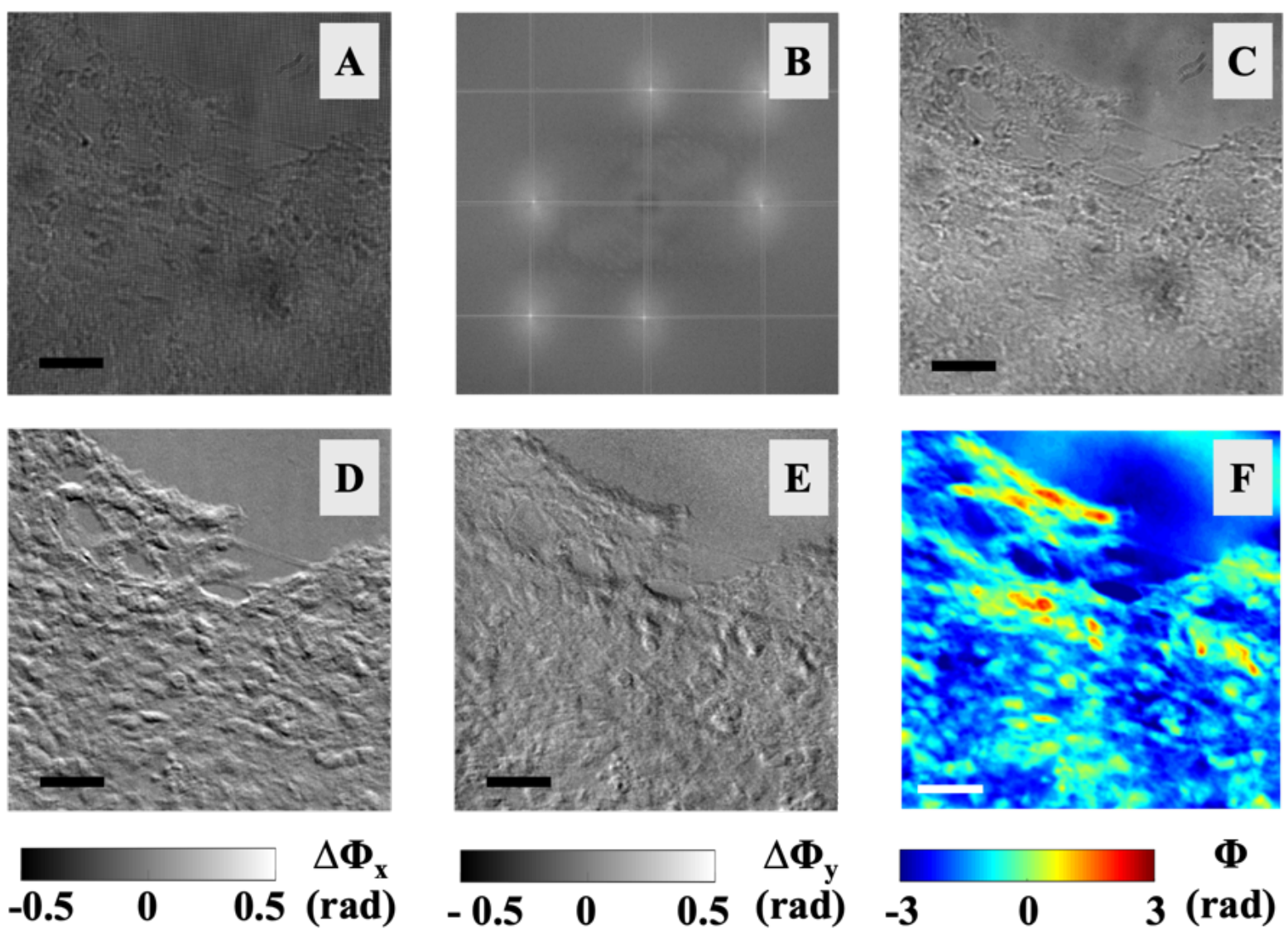

***Supplementary Figure S6 –*** *Thin brain organoid tissue imaging. Raw interferogram (A), Fourier spectrum (B), amplitude (C), gradient phase in X-axis (D), gradient phase in Y-axis (E), integrated phase (F). All scale bars correspond to 20 micrometers.*

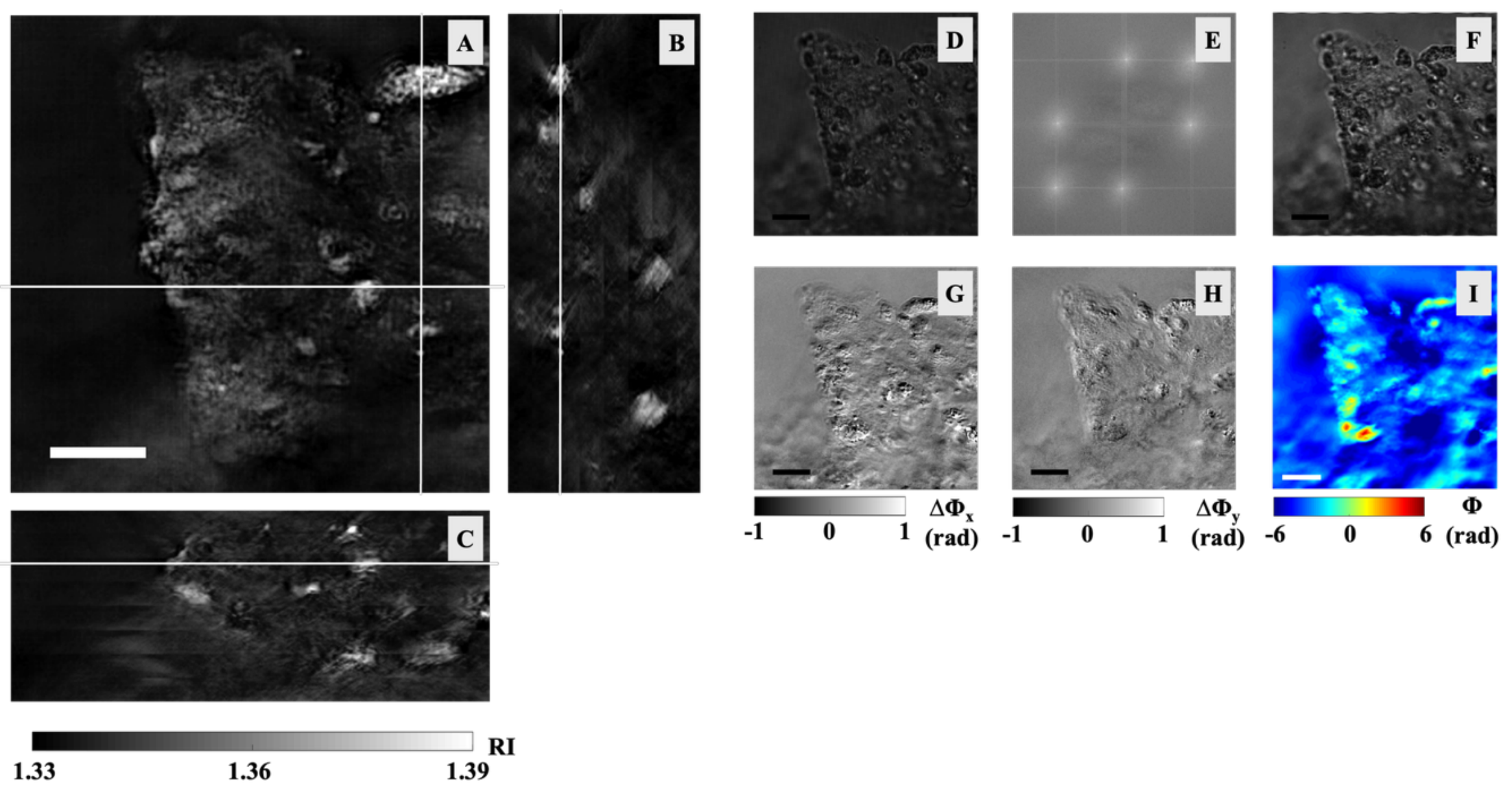

***Supplementary Figure S7 –*** *Imaging results of 80-micron thick brain organoid tissue. X-Y cross-section of RI map (A), X-Z cross-section of RI map (B), Y-Z cross-section of RI map (C), raw interferogram (D), Fourier spectrum (E), amplitude (F), gradient phase in X-axis (G), gradient phase in Y-axis (H), integrated phase (I). All scale bars correspond to 20 micrometers.*

**Brain organoid correlative imaging:**

A more detailed analysis of ROI 3 from Figure 7 is provided in Supplementary Figure S8, which displays the same field of view using six imaging modalities: combined DAPI and MAP2 fluorescence (A), enhanced MAP2 (B), DAPI alone (C), three-channel fluorescence overlay (D), DIC image (E), and RI map (F). A third fluorophore, GFAP (green channel), was also included to label astrocytes. However, astrocytes typically appear later in organoid development (after day 45–60), and the absence of $GFAP^+$ signal in this sample confirms the immature state of the culture. The green channel registered autofluorescence, which, while unintended, aided in multimodal image alignment across channels.

Through correlative analysis of ROI 3, we were able to identify and spatially localize neural stem cells – defined by DAPI-positive, MAP2-negative profiles – across all modalities. These stem cells are highlighted in yellow circles in Supplementary Figure S8. Notably, their morphology and location within RI maps corresponded well with areas of uniform refractive index, further demonstrating the ability of LS-ODT to detect biologically meaningful structures without labeling.

Fluorescent images in Supplementary Figures S8 and Figure 7 were acquired with commercial Delta Vision fluorescence microscope.

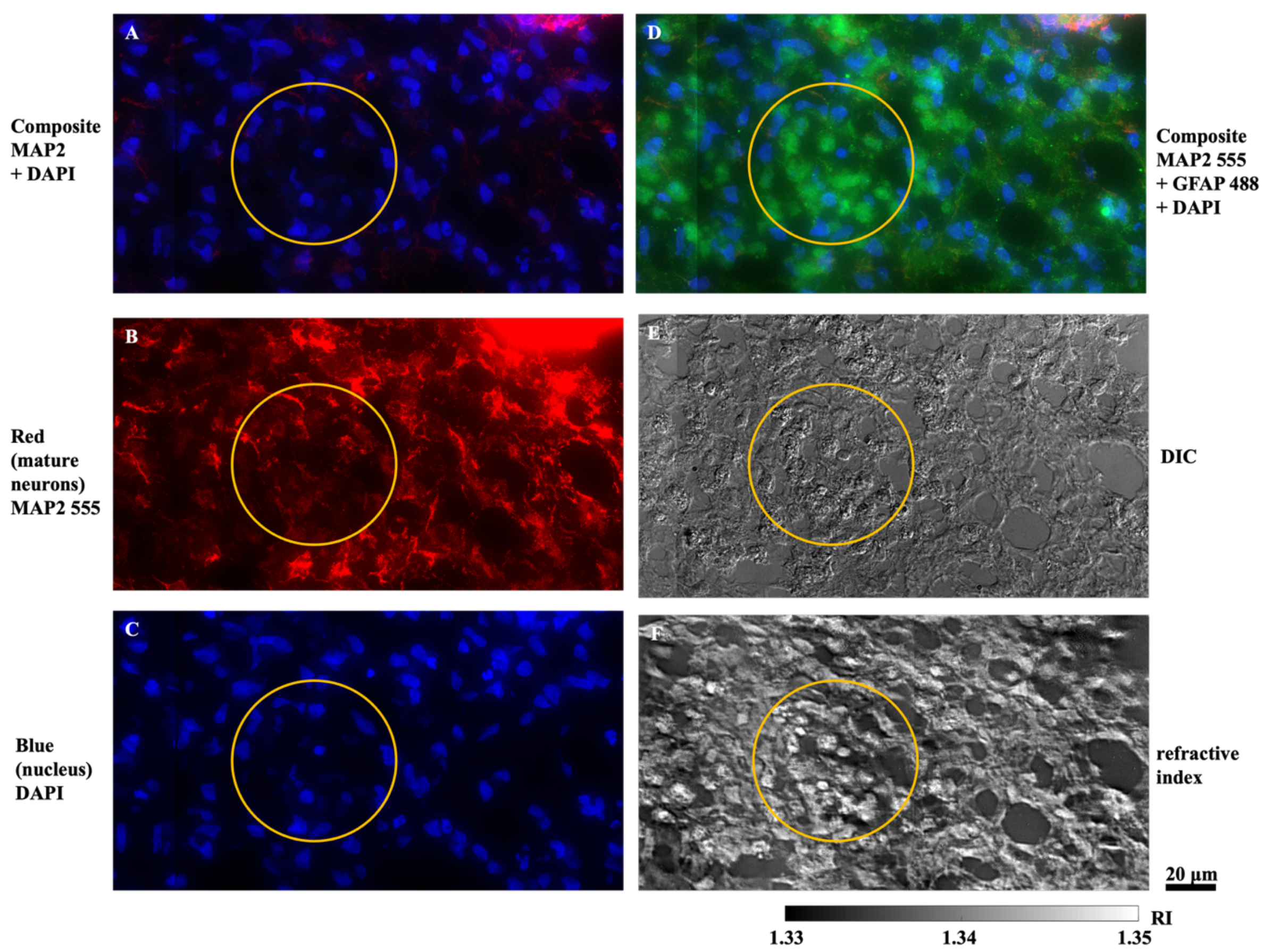

***Supplementary Figure S8 –*** *Correlative imaging of ROI 3 in the brain organoid after 30-day growth. The figure shows the same FOV in 6 different modalities: neurons and nuclei together (A), neurons only (B), nuclei only (C), neurons, nuclei and autofluorescence signal combined (D), DIC image (E) and RI map (F).*